# ICT System Design & Implementation Using Wireless Sensors to Support Elderly In-home Assistance

## Master's Thesis in Embedded and Intelligent Systems

Stefan Plank, Thomas Nowotny & Thomas J. Lampoltshammer

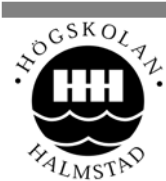

School of Information Science, Computer and Electrical Engineering
Halmstad University

# ICT System Design & Implementation Using Wireless Sensors to Support Elderly In-home Assistance

Master's Thesis in Embedded and Intelligent Systems

August 2011

## Preface

We like to say thank you to all that supported us and without them this thesis would not have become real. First of all we like to state our highest gratitude to our supervisors Edison Pignaton de Freitas and Professor Tony Larsson, as well as Wagner Ourique de Morais. Their experience, advice, help and encouragement have made this thesis possible. Also, we like to thank Josef Höllwerth of K-E-B.com Elektrotechnik GmbH for providing the needed technical equipment regarding the prototype. In addition to that, we like to thank Andreas Unterweger for his profound job as proof-reader of the thesis. Furthermore, we like to give profuse thanks to our families and friends who gave us consolation during this intense and debilitating time.

Stefan Plank, Thomas Nowotny
and Thomas Lampoltshammer

Halmstad University, August 2011

# Abstract

Around the globe the number of older people in relation to the rest is constantly growing. As a result, medical and care facilities cannot handle the growing number of patients. Therefore, elderly in-home assistance gets more attention an importance. Due to issues regarding memory, physical strength and reduced self-assessment, old people face a lot of challenges in accomplishing their activities of daily living. This thesis is meant to address these problems by analysing the required infrastructure of a home-care facility as well as the arising issues regarding used components, especially wireless sensors. After the analysis, a prototype of a home-care system is designed and implemented. Furthermore, the issue of energy consumption of the used wireless sensor node is addressed by modifying the intelligence of the used sensor. After that, the design and components of the prototype used for the energy consumption analysis is explained, together with the programming structure of the sensor nodes used in this thesis. Thereupon, the results are of the simulations are discussed and compared with the authors 'expectations. Finally the overall outcomes of the thesis are analysed and summed up, followed by a short outlook of further possible improvements and developments.

# Contents

# Introduction

## 1.1 Application Area and Motivation

The fact of decreasing birth rate combined with higher life expectation due to improved medical possibilities nowadays leads to an imbalanced relation between young and old people. Therefore, the resulting costs for healthcare increases significantly. Addition- ally, there is lack of qualified and trained personnel. Integration of ICT supports and partly compensates nursing personnel in different areas such as home care for elderly people. Vital signs of people can be monitored and analysed remotely, also known as tele-monitoring.

The introduction of ICT to the fields of health-care and home-care results in a decrease of time and personnel needs, which in return leads to a distinct reduction of expenses. Therefore, a demand for solutions of ICT systems usable in health-care exists.

For the above-mentioned reasons the Austrian company K-E-B.com Elektrotechnik GmbH is interested in an approach of solving the integration problem as well as a prototype implementation.

## 1.2 Problem Studied

Two problems are addressed in this thesis. On the one hand a functional prototype of an elderly in-home assistance system should be developed, that makes it possible to provide surveillance in terms of detecting falls of the patient as a healthcare applications. On the other hand, the energy consumption problem introduced to the system by the use of wireless sensor should be analysed and a possible solution to this problem should be implemented. This will be realized using an approach that modifies the intelligence of the used sensor in order to save energy.

## 1.3 Approach Chosen to Solve the Problem

In order to solve the problem addressed in this thesis, the following approach is going to be used. The thesis will be split into two parts. The first part is going to discuss the design of the elderly in-home assistance system in theory and subsequently present a concrete implementation based on the design requirements determined before. The second part is going to address the energy consumption problem of the used wireless sensor. A concept to solve this issue is introduced and described theoretically. After that, the concept is implemented in the system prototype specified and developed before. In addition to that, simulations are conducted with the modified sensor within the prototype setup to determine the success of the concept and the resulting energy savings.

## 1.4 Thesis Goals and Expected Results

- Design and implement a working prototype for an elderly in-home assistance system
- Provide a working solution to the energy consumption problem of wireless sensors used in the system
- Evaluation of the results and suggestion of further research work for improvements to the developed prototype as well as the energy saving concept

## 1.5 Structure

This thesis consists of six main chapters. In chapter 2, an introduction to the topic of ICT and healthcare is given, providing the necessary interconnection background for the elderly in-home assistance concept. Chapter 3 is then describing the concept of elderly in-home assistance in detail, discussing possible issues regarding the design as well as its implementation. In addition to that, the issue of energy consumption of the used sensor components is addressed as well. From both parts, problem statements are derived which are going to be handled in the prototype. In chapter 4, the design of the prototype as well as its included components are described theoretically and afterwards implemented. Chapter 5 focuses on the energy consumption of a sensitive part of the system that was designed, namely the wireless sensor node which is battery driven. Therefore, the authors introduce a software-based approach to overcome this issue. In addition, the concept is implemented into the prototype and after that simulations are conducted to evaluate the concept and its quality, followed by a discussion of the results. In chapter 6, the conclusion of the work is presented, followed by chapter 7 that suggests further improvements and future work.

## 2  ICT and Healthcare

The population of elderly people[1] in many European countries is increasing. This was proved by via a study by SDI-Research on the example of the population pyramid of Austria[2], see figure 2-1. The Swedish Ministry of Health and Social Affairs published a fact sheet that shows the same developments in Sweden [15]. This development is significant in developing countries and observable around the world [18]. However, European countries exhibit the highest amount of elderly people. Therefore diseases known for elderly people will become more common and healthcare for elderly people will become even more important. Rehabilitation will be moved to private residences to diminish costs by reducing the amount of people staying in hospital beds.

The Swedish government supports home care for elderly and disabled people so that they can live at home [16, p.1]. This would lead to more medical assignments for home-care personnel if people will not stay as long in hospitals as they do now, though the number of qualified personnel has not been raised as it would be necessary to cover the increased demand [16, p.1]. To receive home care from a municipality in Sweden a citizen must apply for aid. A caseworker will examine the application and decide on the goal, the amount of care needed and define what kind of care is required.

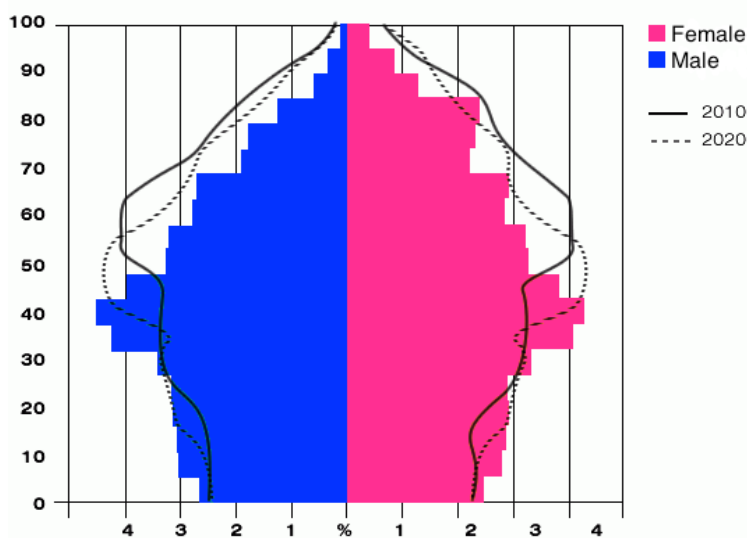

**Figure 2-1: Demographic Pyramid of Austria**

The home care personnel will then visit the person and help them in the determined way [16, p.8].

The majority of the personnel are staff nurses[3]. Women mostly carry out a care-related profession. It is often low paid and has no high status in society. This leads to a difficult situation regarding the recruitment for new personnel. Care personnel often lack the adequate education. The government expects a positive influence on the status of the profession due to research and integration of technology into these jobs [16, p.1-2].

---

[1] The term "elderly people" refers to persons that are 65 years or older
http://www.who.int/healthinfo/survey/ageingdefnolder/en/index.html
[2] http://www.sdi-research.at/lexikon/alterspyramide.html
[3] http://www.lasvegassun.com/news/2008/may/27/man-woman-nurse-engineer/

Information Technology (IT) solutions for home care are already available, though mostly not in use in this area. For example, tele-cardiology could be used to monitor vital signs[4]. The use of Information and Communication Technology (ICT) will be obligatory in the future to provide the possibility for elderly to stay at home. The work should be more effective and performance evaluations of the services provided could be done more efficiently.

The improvements reach beyond the use if ICT could also increase the job satisfaction of the care personnel itself and therefore improve the work situation and the image of the profession [16, p.2].

ICT has a large potential as hospitals already use advanced IT systems for analysis and embedded technology for operations [27, p.9]. Administrative tools for coordination and communication are in use as well. However, these systems are not commonly used in elderly care. Hence, the required communication between different systems is not possible due to missing standards in that area. Moreover, in order to benefit from ICT organizations have to adapt to the system and train their employees on how to use it properly [27, p.9]. Care assistants have been working with some kind of technology for a long time as they use ergonomic and household technology like lifts to move people from or into a bed [16, p.4]. IT Systems are often used as administrative tools for timetables and organizational issues, though they are rarely used in operative work regarding mobile solutions like sensor technology or remote monitoring.

The development and implementation of ICT in home care, focused on how to improve the services provided by the care assistant and to reduce costs, make long term planning necessary [16, p.7]. In addition, political support is vital - there have been projects, though not in a large enough scale to connect multiple organizations [16, p.7]. Another issue regarding the implementation and development of home care systems is that the home care personnel have mostly no influence regarding the processes or decisions related to the bought equipment. Medical informatics systems are built to provide medical expertise; knowledge and supportive work with patients is often not seen as important as it should be [16, p.41].

Furthermore, the communication between all involved parties could be improved. Patient related information like medication or treatment plans for rehabilitation are not available for care assistants [27, p20]. To improve the communication between the governments, the hospitals and the home care organizations a common IT system or at least a standardized interface is needed. As a result the integration of ICT systems would be much more useful in home care. Moreover, the integration involves different parties like the health care personnel, technology suppliers, the owners of residences and the government [27, p.1].

---

[4] http://www.vitaphone.de/en/home.html

However, integration alone is not enough. The personnel needs to be advised how to work with it efficiently and procedures have to be changed to adapt to the system [27, p.9].

Though, integration is yet not common; only hospitals use advanced technology in the medical area. Nowadays, IT systems are well known and have been used in many different areas of application. Therefore, technical difficulties are not the cause to delay the development [27, p15]. Though, the impact of these systems in the area of home care is not decreasing costs and improving the quality of the service. In order to improve the situation, different organizations would need to work together, which fails due to uncertainties regarding payment responsibilities. As a result, the development is delayed significantly [27, p.19].

The advantages of ICT in home care are for elderly people to stay at home longer in their familiar surroundings. Their home needs to be equipped with ICT networks and most houses are connected or could easily be connected to the Internet. Sensors need to be installed, which would cause costs in the beginning. The health condition of patients could be improved. They will not need to go to doctors as often for checks if they do not want to due to their vital signs could be monitored remotely. Permanent monitoring would make it possible to diagnose possible issues earlier and improve treatments for the patient. Overall, this leads to a more efficient way of using health care combined with reduced costs, seen over a long-time period. Hospitals and healthcare centers need the possibility to monitor the sensors of all people in their area. Furthermore, the home care personnel need to learn to cope with ICT Systems.

Overall, the integration of ICT leads to a high investment at the beginning, which involves all parties working in healthcare including hospitals and the government, although it increases the quality and safety of home and healthcare for everybody [27, p.25].

## 3   Elderly In-home Assistance

As mentioned in the chapter before, ICT and elderly in-home assistance are getting wider importance and attention. This chapter will discuss technical related ICT system issues as well as upcoming aspects due to used components in the system for elderly in-home assistance. Furthermore, problem statements will be developed regarding the identified issues.

### 3.1   System Design

Elderly in-home assistance (EHA) aims at the relocation of medical and health services from medical institutions to the patients' home. These services range from nursing personnel for chronic illnesses or illnesses which can be treated in an ambulant way up to support elderly people to live their daily life in their own homes. Included in this daily routine are activities of daily living (ADLs)[5] such activities include self- care (e.g. eating, dressing, body cleaning), work and leisure. The difficulties here lie in the fact that elderly people tend to display reduced capabilities regarding memory, movements and sagacity[6]. This can lead to dangerous situations like the scenario of a forgotten cooktop after dinner, falls or body injuries due to heavy lifting or climbing on furnishings. Therefore, EHA has to provide functionalities for monitoring the patient as well as the environment the patient is living in and report any possible source of danger or accident to the caregivers or responsible persons like family doctor or relatives.

Upcoming challenges regarding the above specified requirements are that the patient wants to remain independent and self-sustaining as long as possible. Hefty sensors, wires, or complicated user interfaces to use EHA services are barriers regarding the patients' ability to roam freely inside her home. Hence it is important to consider these aspects when designing an EHA system concept. The next section will show possible components and services of an EHA system.

---

[5] http://www.medterms.com/script/main/art.asp?article
[6] http://www.altenpflegeschueler.de/psychologie-soziologie/allgemeine-Wahrnehmung-2.php

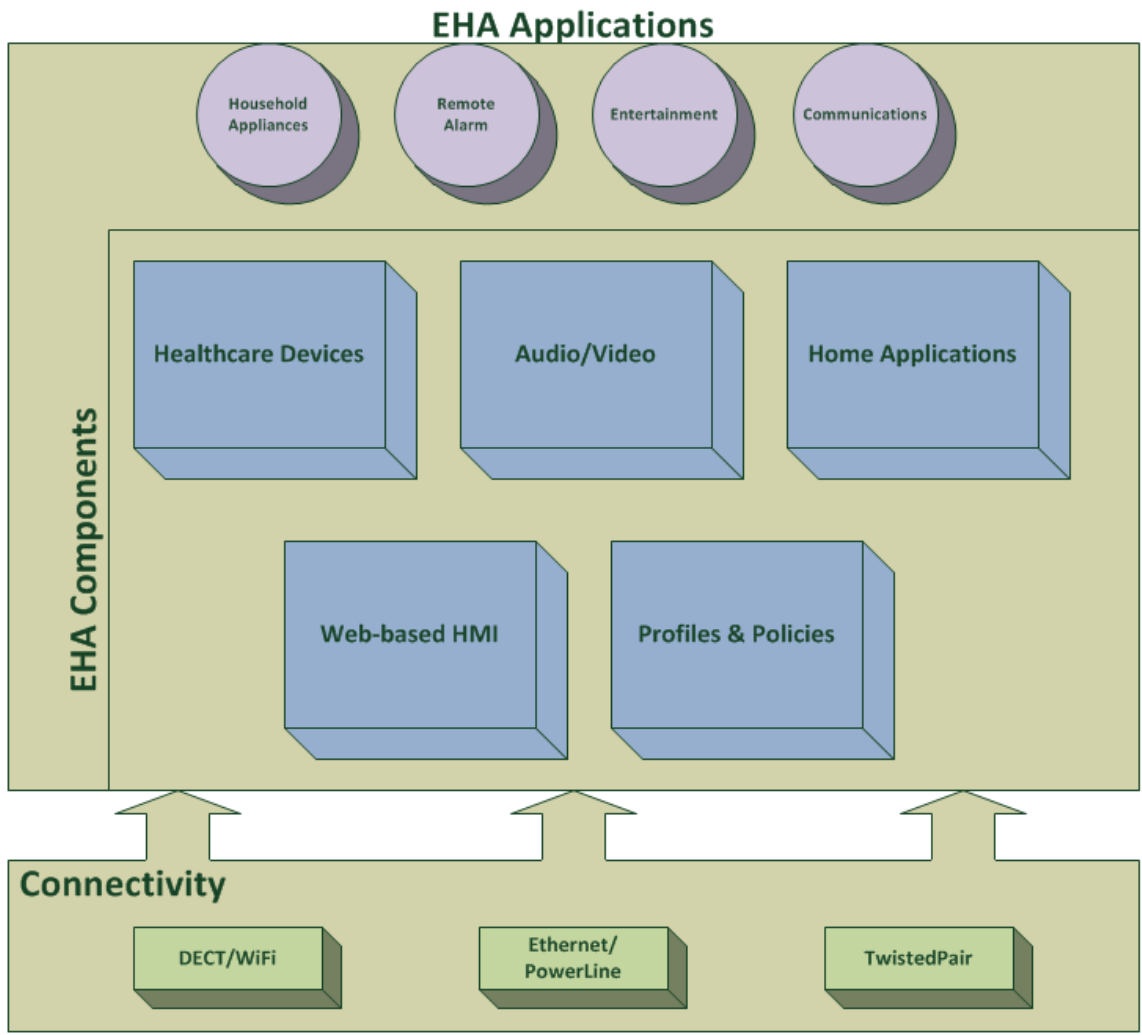

**Figure 3-1: Components of an Elderly In-home
Assistance System, adapted from [42]**

As can be seen in figure 3-1, the three main blocks in an EHA system are
connectivity, the components and the applications. In order to provide connectivity
inside the home as well to remote locations several technologies may be used, may it
be wired or wireless. On the side of wireless technologies there are DECT and WIFI
solutions (GPRS, 802.11x, Bluetooth). On the side of wired possibilities dedicated
infrastructures like Ethernet or twisted pair may be an option as well as shared
infrastructure like PowerLine technologies.

The components of an EHA system can be split into four main parts as there are human machine interfaces, healthcare services and sensor APIs, implicit network operation component and remote monitoring and maintenance components [42]:

➢ **Human Machine Interfaces** - On the one hand, the user interfaces have to be easily accessible and self-explaining if they are used by the patients directly, on the other hand they have to be advanced enough for the trained care personnel and they have to be reachable via external connections, so remote access is important as well.

➢ **Healthcare devices and sensor APIs** - The server infrastructure must be able to communicate and interact with the integrated healthcare devices and sensor systems. Therefore, suitable APIs and middleware solutions have to be provided.

➢ **The implicit network operation component** - This component interacts in a proactive way by identifying the user's needs and preferences depending on his/her identity. Furthermore, it provides interaction with remote health centres via text, voice or video.

➢ **Remote monitoring and maintenance component** - These functionalities pro- vide remote monitoring capabilities in order to observe several in-home assistance components. Therefore, it is possible to identify possible failure and dropouts of components early enough. This check can include body functions of patients via sensor devices as well.

The major focus in this thesis is on healthcare applications related to elderly in-home assistance. These applications can be categorized into four groups [1]:

➢ **Activities of daily living monitoring** - Here, the system tries to track the daily activities of the patient like watching TV, sleeping, eating and detect abnormal situations.

➢ **Fall and movement detection** - These applications are focused on physical condition surveillance of the patient. Especially elderly patients that are recovering from operations are prone to falls and as a result can sustain heavy wounds or even death.

➢ **Location tracking** - If patients are cognitively impaired, these applications can help to keep them save and support them in order to still live independently.

➢ **Medical status monitoring** - Applications settled in this area make heavy use of medical and environmental sensors to obtain the health status of the patient. Examples here would be blood pressure, heart rate and oxygen saturation.

Regarding the above- m e n t i o n e d applications for healthcare, the authors like to focus on the fall and movement detection in their work. In the upcoming section, related work will be presented that is dedicated to elderly in-home assistant systems as well as healthcare applications regarding fall and movement detection.

### 3.1.1 Related Work

The design of EHA systems is an active scientific research field. In the work of Zhou et al. [54] an approach towards a case-driven ambience intelligence system was pursued. This system and the included technologies are targeting the support of ADLs by combining multiple sensor information and, based on them, make decisions and act accordingly. Therefore, an information model was designed based on predefined cases and transformed to an embedded platform usable for EHA in a smart home.

Another example can be found in the work of Yao et al. [51]. They experimented with wearable devices and wireless sensor networks to improve plug-and-play interoperability between infrastructure and wearable devices. They concluded that it is not possible to get rid of human interaction but sensor devices and fitting infrastructure can indeed heavily support the patient if human interaction and technical environment are well combined.

Purwar et al. [30] have done an interesting work in the application area of fall and movement detection by the use of a real-time triaxial accelerometer. They sensor used gathers the data from the accelerometer and transfers it for further processing to a central node. They achieved an accuracy of 81% regarding a correct detection. They point out that if the sensor is worn with a higher altitude to the ground (for example at the chest), the results are far better than if it was worn at the wrist.

Also settled in the application area of fall detection is the work of Wang et al. [44]. This version of accelerometer works with a head-level attached sensor, which is able to distinguish abnormal situations like a fall from ADLs. Via a self-developed algorithm eight kinds of falling situations as well as seven daily activities can be differentiated. The method behind this algorithm is based upon taking the difference between the initial time of the body hitting the ground and the time when the body of the patient is in rest.

Another research work by Leijdekkers et al. [21] proposed the concept of measuring fast accelerations and determines if a fall has happened or is not related to the movements after the acceleration has occurred. However, this approach has been marked as unreliable due to the fact that a patient can move after a fall due to pain.

### 3.1.2 Problem Statement

In conclusion, it can be stated that the main problems regarding the design of an elderly in-home assistant system are to support the patient without withdrawing his or her independency while at the same time provide constant surveillance and support of ADLs. It is also important to include smart decision capabilities into the system so that external personnel are only consulted if really necessary. By doing so, reduction of accruing costs to a minimum gets possible.

## 3.2   Sensor Related Energy Consumption

Energy consumption is not only an important topic regarding the whole system design, but also has a huge impact in a financial aspect. The energy costs for a smart home have to be paid and in addition to that the battery exchange of mobile device is not only time consuming but also produce more waste and extra costs. According to an experimental research study [31] with several non-experimental sensors it turned out for the communication portion of the sensor node being the most energy consuming part. The energy expenditure by processing data on the sensor node was marginal. This can be exemplified by the fact that processing several bits consumes less energy than the transmission of a single bit [41]. In order to put these findings into numbers it can be declared that communication of a sensor node occupies up to 80% of the total energy consumption, while processing the data and sensing functionalities only require 20% [24], as can be seen in figure 3-2.

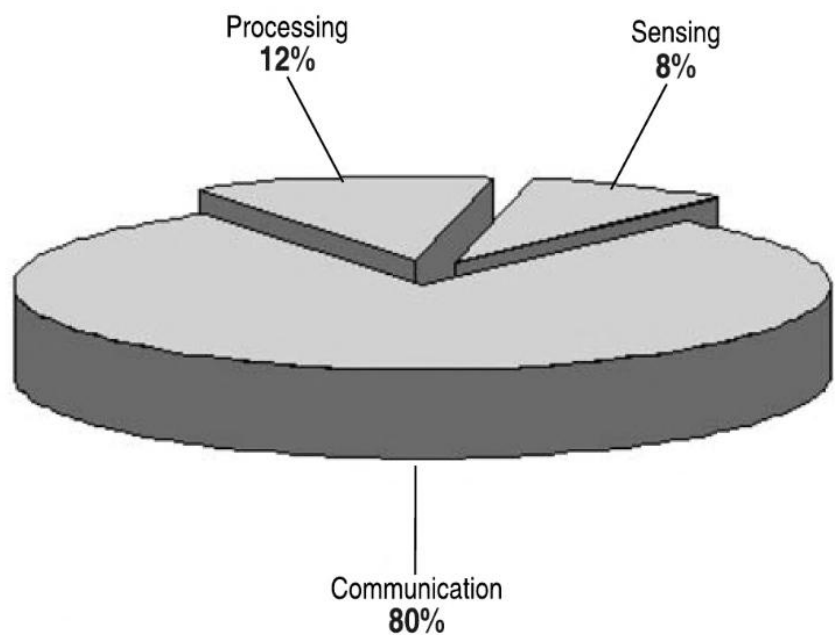

**Figure 3-2: Energy Consumption of a Sensor Node**

Two general approaches can be identified in order to counter this problem. One of them is energy scavenging or energy harvesting [46]. The concept is based on gaining energy from the surrounding environment. Examples therefore can be found in solar energy, mechanical energy like vibrations or thermal energy, as well as gaining energy from electro-magnetic fields. A major challenge in this research field is to provide an energy supply without any interruption, which requires at the moment a battery as a buffer. By using energy scavenging techniques it is possible to use a green form of energy and to support the sensor node with a theoretically unlimited energy supply [23]. The second approach is based on methods for energy conservation. The concept behind this term is to reduce energy consumption in the designated device in all components.

Three forms of energy drain can be identified, as there are active current consumption, standby current consumption (sleep mode) and off-mode current consumption. The first kind of drain can be described as the sum of static energy consumption and dynamic energy consumption caused by switched or clocked device parts. The second form is nearly only based on static energy consumption due to dynamical drains being removed by stopping switched and clocked components. The third form of energy drain presented is caused by the sub-threshold leakage of transistors due to the power supply still be connected to the chip event if it is not active. In order to take actions against these forms of energy drain there exist several techniques covering all layers of a device starting with silicon IP, over SOC design up to dedicated system software. The company Texas Instruments (TI) calls its approach SmartReflex Technology [6]. As it can be seen in figure 3-3, the approach consists out of three layers, each of them covering a specific technical aspect of how to preserve energy in wireless devices.

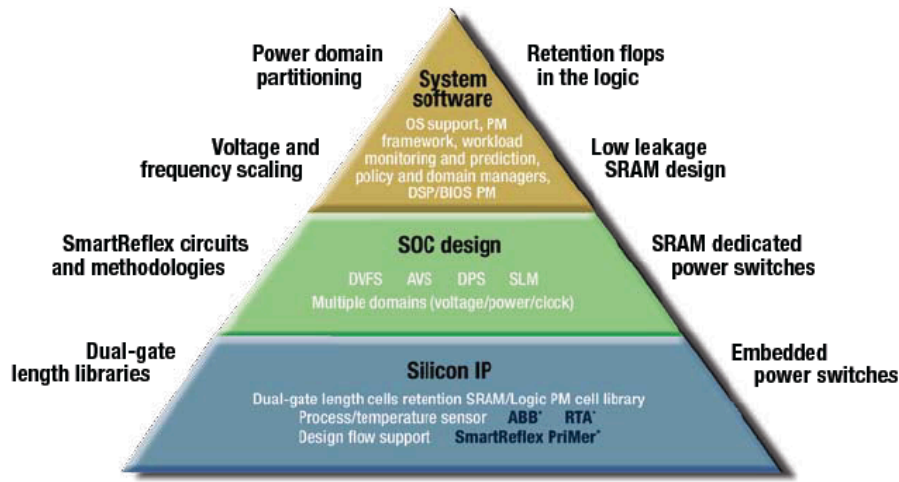

**Figure 3-3: Smart Reflex Technology Pyramid**

### 3.2.1 Related Work

A lot of research is currently conducted related to the issue of energy consumption of wireless sensor devices. One example is the paper by Zhang et al. [53]. It discusses the concept of using two transmission channels instead of one based on the hypothesis that it is often not necessary to keep a transmission channel alive continuously and that it is sufficient to trigger data transmissions off the sensor node on-demand only. The introduced second channel is dedicated to surveillance purposes. Thus, it enables the sensor node to provide transmission on-demand capabilities combined with zero energy consumption during idle time. In respect to the second channel, only a receiving unit is required. Such a receiving unit can be realized by installing a RFID module on the sensor node. In return, the RFID offers benefits in terms of instant response as well as energy scavenging and no idle energy consumption while listening. When being woken up the RFID module activates the regular transmission module and the gathered data of the sensor can be transmitted via the first channel.

Mondal et al. gave another example in [26]. The proposal of this work is to reduce the payload of the transmissions via the application of compression techniques like LPC. In order to prove their point, they conducted a comparison between this method and others as well as plain transmissions. The proposition of this paper also aims at the reduction of transmissions. Though not based on compression techniques, the sensor node's intelligence is altered to reduce the transmission payload instead.

However, not only energy consumption is a possibility. As mentioned before, there lies great potential in energy harvesting as well. An on-going development shows interest not only in green energy by solar energy, but rather in body related energy sources as well. On example is the work of Renaud et al. [32] or Lauterbach et al. [20] and Leonov et al. [22] that are focusing on gaining energy by exploiting the motion of the human body or its thermal energy.

### 3.2.2 Problem Statement

In conclusion, it can be testified that the problem of energy consumption has to be addressed by either energy harvesting technologies or energy conservation approaches. Due to the fact that the hardware platform is often not modifiable, software-based approaches promise compatibility as well as potential regarding energy conservation.

# 4 System Design and Results

This section discusses the components and architecture of the prototype, while the section after that deals with the technical implementation. The sensor nodes' intelligence and its implementation are described in the last two sections.

## 4.1 System Concept

The prototype fulfils two main tasks. First, a surveillance network including sensors and a camera must be set up in order to guarantee monitoring of a person's vital signs and movement. This includes detection of certain patterns, in this specific case a person's movement and fall. Additionally, the prototype provides a centralized platform that handles sensor and network traffic and enables management functionalities such as camera control. Furthermore, emergency notifications are sent via the network to remote devices. Figure 4-1 shows the architecture of the used prototype in a general way.

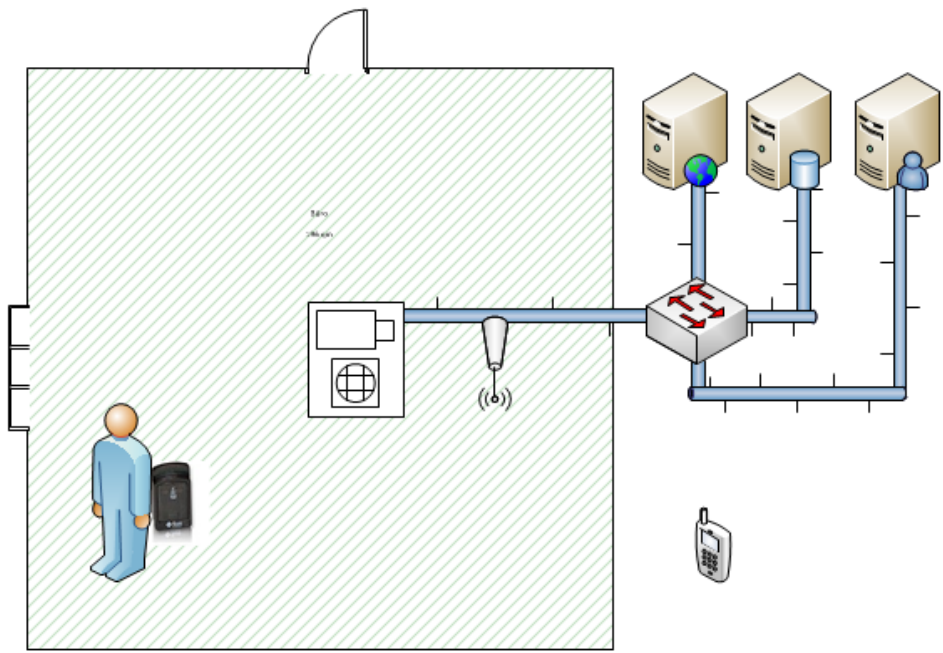

**Figure 4-1: Prototype System Concept**

According to figure 4-1, several key systems and devices where chosen to accomplish the given tasks described before. The prototype system can be divided into three main parts: data gathering and transmission (DG&T), monitoring and surveillance (M&S) as well as management, control and storage (MCS). Figure 4.2 depicts how these components are interconnected and their data-flow.

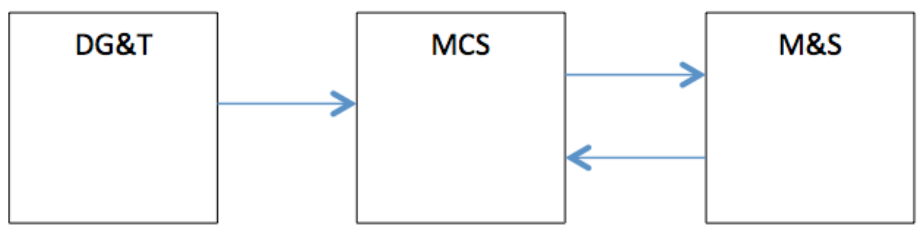

**Figure 4-2: Proto 2**

Sensor nodes and base stations as well as wireless transmission are part of DG&T block.

Cameras and mobile notification devices are located in the M&S part. MCS contains a web server to handle incoming sensor data from the base station. In addition, camera images or streams are displayed in this block using a management tool and notifications are sent to remote devices via the SIP protocol. A database server enables storage of either sensor data or camera images. Collected sensor data is sent one-way to the management and control block where it is processed.

A bidirectional communication is established between the management, control and storage part and the monitoring and surveillance part to allow outgoing emergency notifications as well as incoming camera data. Each component used in the prototype is described in detail below:

➢ **Sensor Node** - Sensor nodes are used to gather any kind of information, e.g. heart rate and movement of persons (or patients) or room temperature. Depending on required data either mobile sensor nodes e.g. carried by a person to measure vital signs or stationary sensors are used to collect data. This allows a setup of sensor networks to measure different physical aspects of rooms and patients simultaneously. Measured data is sent to one or more base stations using wireless communication for mobile sensors or wired communication for stationary sensors.

➢ **Base Station** - The base station is used to receive data sent by sensor nodes or a sensor network. Therefore, its technological aspects depend on the sensor nodes in use. If mobile sensors are used, the base station must be able to handle wireless communication. The base station is responsible for routing data and information collected by sensors or sensor networks to other connected networks or servers where data is further processed.

➢ **Camera** - Cameras are installed in different rooms as stationary surveillance devices. Each room equipped with either stationary or mobile sensors should be monitored by at least one camera. This ensures visual coverage of areas where sensor nodes are present. Present cameras allow capturing situations when sensor node data indicates an emergency. In some cases it is not necessary to monitor a certain area the entire time. Camera images or streams can be triggered by certain sensor data patterns and emergencies to allow privacy as long as possible. These systems could be used in a retirement centre for example.

➢ **Session Initiation Protocol (SIP) Server** - A SIP server is used to allow Voice over Internet Protocol (VoIP) communication between prototype components. SIP requires an IP-based network to function. Existing wiring inside a building can be used as well as the Internet to connect remote facilities. Communication terminals can be installed inside rooms where sensor nodes are operational to allow communication from VoIP phones or VoIP mobile phone applications to a room's communication terminal. This connection is established manually by dialling the terminals number or each time a sensor data pattern triggers an emergency event.

➢ **Mobile Device** - Any device capable of using SIP and VoIP serves as a suitable solution for the prototype. Several smart phones provide the opportunity to install SIP applications that allow SIP communication when connected to a wireless local area network (WLAN).

➢     **Web Server** - The web server used in this prototype receives and processes collected sensor data. The server also triggers certain events depending on sensor data values. An example is an emergency notification using VoIP functionality of the SIP server if a sensor measurement exceeds a pre-set threshold value or specific sensor data is received. Furthermore, it allows analysis and processing of collected data that is stored in a database server afterwards.

➢     **Database Server** - A database server is used as the prototypes storage device. It uses a relational database model to store any kind of data received by the web server, e.g. sensor measurement values and camera images. The web server and the management tool are components that can store data as well as access-required information from the database server using SQL queries.

➢     **Management Tool** - A management application provides all necessary functionalities to control the web server. It also handles displaying camera images or streams and storage or export options for sensor measurements. Mainly it acts as the human-computer-interface (HCI).

➢     **Network** - A suitable network is necessary to connect all mentioned components above. It is required to support TCP/IP in order to allow SIP and HTTP to function properly. An Internet connection is not required but would allow remote access to the servers or distributed server architecture. Furthermore, VoIP connections can be established independent of the remote devices position.

## 4.2 System Implementation

In order to keep the prototype simple, several prototype components are combined and not implemented individually. Web server and management tool are embedded into one application and the SIP servers as well as the database server are installed on a single workstation which also runs the application mentioned before.

### 4.2.1 Sensor Setup

Sun's (Sun Microsystems was taken over by Oracle in 2009. Despite that fact, the authors will stick with the original name Sun instead the official name Sun Oracle) Small Programmable Object Technology (SPOTs) represents tiny, battery-driven de- vices equipped with wireless communication technology in order to provide a new, future-oriented prototype and development platform for Java-based network programming. Application scenarios are settled in the fields of environmental surveillance, robotics, asset tracking and e-health for example. The core of the SPOTs is represented via a Java virtual machine (VM) which was especially designed to meet the SPOTs' demands. An own section is dedicated to this VM later on.

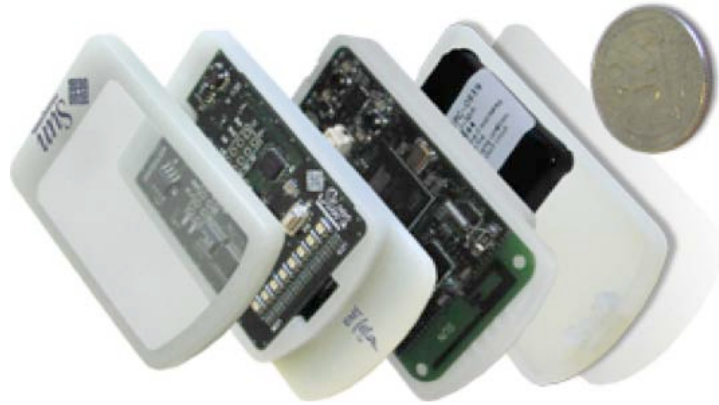

**Figure 4-3: SPOT and Each of Its Layers, adapted from [17]**

The SPOT is built out of several stacks that in total provide sensors and actuators for a huge variety of applications. On the boards in the single stacks there are accelerometers, temperature sensors, LEDs, buttons, light detectors and I/O pins for attaching external sources as well. Depending on the duty cycling used, the SPOT can maintain operative for months by one single charge of its battery that is rechargeable.

The big advantage in comparison to other embedded development platforms is the fact that it is not necessary to learn new and unfamiliar programming languages or use complicated, sometimes expensive tools. Also debugging is often a major issue. For the SPOTs, standard Integrated Development Platforms (IDE) like NetBeans or Eclipse can be used due to the use of Java [39].

### 4.2.1.1 Squawk Virtual Machine

Squawk is a VM based upon the Java Micro Edition (Java ME) and is dedicated to embedded systems and programming of small devices. The standard VMs for Java are coded in low-level programming languages like C or C++ and on the deepest level in assembler. The main difference to these VMs is the fact that the core for Squawk is nearly up 100% Java-based and therefore it is out of the group of meta-circular interpreters. For this reasons Squawk VM is the optimal solution in this case for a perfect integration of Java resources like interfaces, threads or objects.

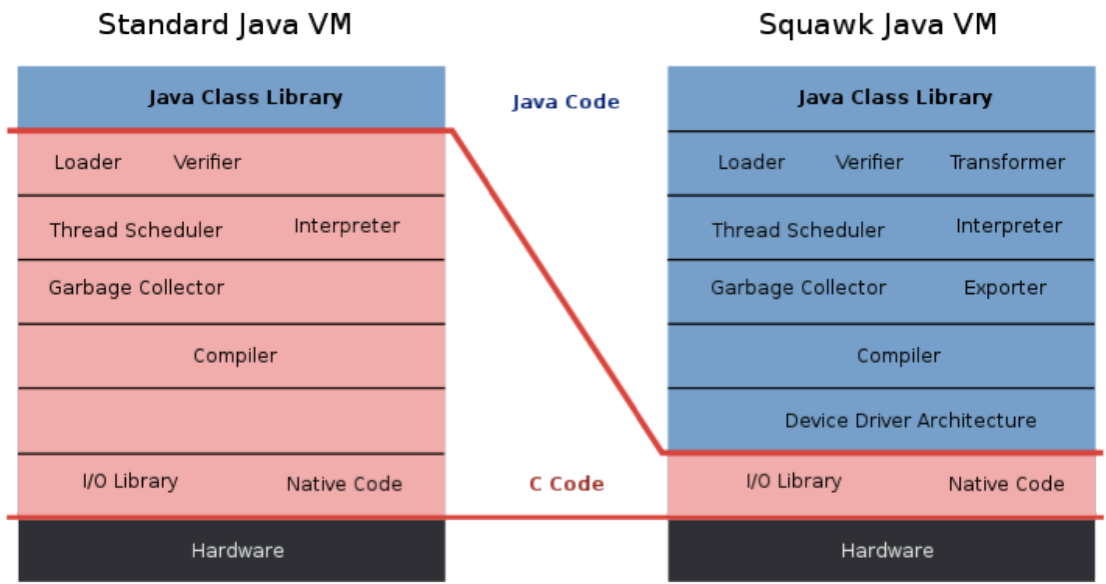

**Figure 4-4: Standard Java VM vs. Squawk Java VM,
adapted from [49]**

Squawk comes with a very small footprint regarding memory demands and minimal external dependencies. Inside the VM all applications are treated, as objects and therefore are completely isolated from other applications running on the SPOT inside the VM at the same time. Some of the most important features are e.g. the capability of On-Device Dynamic Loading and Linking paired with On-Device Verification which means that program components are dynamically loaded and linked on the SPOT as well as verified providing same reliable results as the Connected Limited Device Configuration (CLDC).

Another advantage comes with the Exact Garbage Collection. The VM uses the principals of garbage collection akin to the correspondent on desktop computers. Its special ability lies in the knowledge of all references and their whereabouts. Therefore, it can trace and collect all of them completely. Also, its EEPROM-aware memory system brings stability and performance by dividing the existing memory into three main spaces namely ROM, RAM and electrically EEP- ROM.

In addition to the above-mentioned feature, the VM profits from the extremely compact class-file representation used. The Squawk VM converts class-files into a much more compact form in order to provide the above mentioned profits. This change leads to a reduction regarding off-device storage and transmission demands, together with on-device loading requirements [37].

*4.2.1.2 Solarium*

The Solarium suite comes along with the Sun SPOT package and provides an environment for remote management of networks built by SPOTs. For Solarium being a Java application itself, it is also platform independent. Managing SPOTs via Solarium makes it possible to administrate applications running on SPOTs during runtime. Solarium is capable of detecting SPOTs either via the USB interface or via radio connection.

Once detected, the suite can be used to deploy and un-deploy software on the spots, together with basic commands like start, pause, resume and stop regarding the deployed applications. While connected, it is also possible to check the status of the SPOTs in terms of battery level or memory capacity. Special views like the radio view or the deployment view make it easier to view and analyse data gathered from the SPOTs.

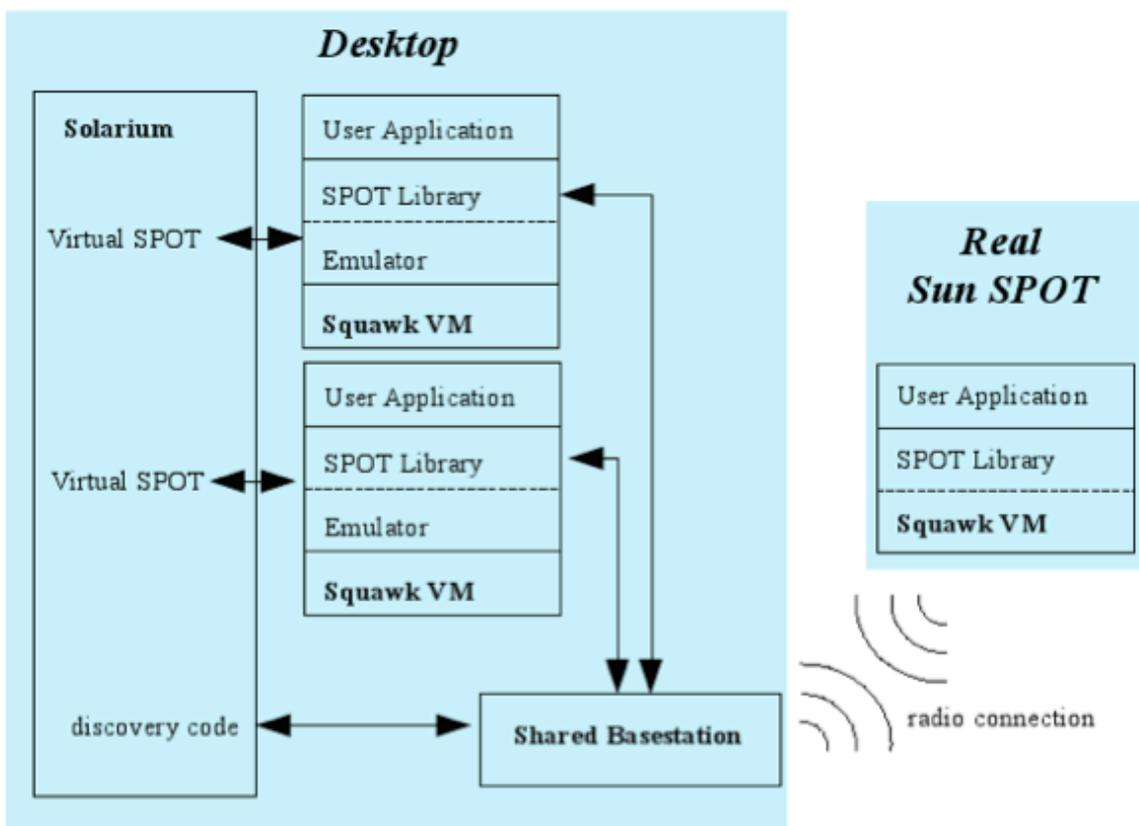

**Figure 4-5: Solarium Network Setup, adapted from [38]**

In the deployment view it is possible to specify a certain setting of a group of SPOTs forming a network structure. One can set the specific application on each SPOT and by a one-click action all SPOTs are configured accordingly at the same time.

Also dependencies and states of the SPOTs can be monitored. Another important feature is the emulation of a SPOT. Emulated SPOTs act like real SPOTs. Java programs can be deployed, ran and debugged as well. It is also possible to form a heterogeneous network structure with emulated as well as with real SPOTs [38].

### 4.2.2 Base Station

A base station is a computer that handles all wireless sensor connections and communication. To provide multi-sensor connection, each sensor is connected to a different port on the base station. The base station acts as a converter for wireless communication to wired communication (TCP/IP). Wireless data packets are therefore parsed and valid information is extracted which is then sent over the IP network using HTTP post commands.

### 4.2.3 Camera Setup

As the main camera a Mobotix Q24 is used. It provides easy setup and access via HTTP, so it can be configured using any web browser. The camera is fully accessible using defined URL commands. The lens enables a full panorama view when mounted on the ceiling. The resolution ranges from 160x120 pixels to 2048x1536 pixels and it only requires a light intensity of 1 Lux when acquiring pictures at 60Hz or 0.05 Lux at 1Hz when working in full colour mode. Black and white colour mode only 0.1 or 0.005 Lux are required [25]. The camera also provides loudspeakers and a microphone as well as an integrated VoIP SIP client for voice communication from and to the camera.

The SIP client can be configured via web interface and a SIP server is necessary for further SIP calls. These camera functionalities replace a SIP transmission terminal in each room where this camera is installed.

### 4.2.4 Web Server and Application Setup

The web server application is implemented in C# using Windows Forms. The application uses the .NET Framework 4 which provides all necessary functionality and components. The implementation can be divided into four main parts:

- Web server
- Camera control
- Database management and control
- Visualization

The web server part provides methods and functionalities for establishing HTTP connections from and to HTTP sources, such as base stations. It also enables reading of HTTP requests such as sensor values sent via HTTP post commands. The web server also implements a self-testing functionality as well as logging and export functionalities for received HTTP commands. The camera control part covers full functionality of handling camera commands, e.g. image acquiring as a live stream, error handling and SIP emergency calls.

It uses Web Request and Web Response classes included in the .NET Framework 4 to communicate with the camera and to execute commands directly on the camera. Database management and control is implemented using the .NET Framework 4 SQL- Client classes. This part is responsible for establishing database connections (open, close) and error handling. Functionalities for storing information into the database as well as reading database entries are implemented here. This part has direct relation to the visualization part where database information is displayed. The visualization part provides display functionalities for database entries using the .NET Chart Control. It therefore uses the database management functions to extract data from the database and displays it using different types of graphs, e.g. lines, bars, dots and so on.

## 4.2.5 Web Server

The web server is controlled by interaction with the application form. The web server tab contains buttons to start and stop the web server as well as a self-test button. When the start button is pressed, a label indicates whether the start was successful or not. Pressing the same button that was used to start the server can stop the server. If a self-test is run, another label shows if the test was successful or not. Furthermore, the web server tab contains a text box where incoming post requests are displayed. Each request is time-stamped and then displayed. This log contains the IP address and port from where the request was made (remote system, base station) and a destination IP address and port (web server) as well as detailed information on the transferred data. The connection log may be exported using the export button or cleared using the clear button.

The main functionalities are encapsulated into a Web Listener class. This class can be instantiated by calling the constructor of this class and an optional port parameter may be passed. Without the parameter passed, the Web Listern class is configured to listen on port 8080. The first member of this class is a form parameter that can be set on demand. All the logging is done to this form. This member should be set to the main form after having the Web Listener class instantiated. The second member is an Http Listener class found in the System.Net namespace.

This class provides full functionality for listening on specific ports for defined IP addresses. The listening procedure blocks all application functionalities. A solution to this problem is to create a separate thread to handle the listening. This leads to the third member, as it is a thread that is responsible to solve the blocking problem. The fourth and fifth members are the port to listen on a Boolean variable that indicates whether the Web Listener class is currently listening or not. Furthermore, an ENUM for self-testing is added which can be 'SUCCESS' or 'FAILED'. The Web Listener class also encapsulates a nested Worker class that handles all necessary HTTP request processing. This class is described later on. The class constructor sets the port and the Boolean variable to false; it also instantiates a new Http Listener class. It is important to define which IP address the Http Lister class should listen on. This is done as follows:

*p = port;*
*l = new HttpListener();*
*string s = "http://+:" + p.ToString() + "/";*
*l.Prefixes.Add(s);*

The '+' pattern in the string which is added to the prefixes of the Http Listener instance indicates to listen to any IP address which sends HTTP requests. The main functions of the Web Listener class are Start (), Stop (), Listen (), Test () and Write2Log () where the Listen () and Write2Log () function are not accessible from outside the class. The Start () function simply creates a new thread that executes the Listen () function and sets the Boolean variable to true. The Stop () function stops the Http Listener that was started in the Listen () function and sets the Boolean variable back to false.

The Listen () function starts the Http Listener instance by calling its Start () function and afterwards accepting incoming HTTP requests by calling the GetContext () function. This function is the blocking part for which the thread was created. It waits for the next context to arrive. Each accepted context from the GetContext () is the passed on to the nested Worker class and an additional thread is created to handle the processing.

The Test () function simply creates a HTTP post request that is sent to localhost. Depending on the reachability of the server, it returns 'SUCCESS' or 'FAILED' as the test result.

The Write2Log () function is called each time a context has been passed to the Worker class and if a form was set, it writes information to the log text box using the appropriate delegate defined in the form. The Worker class only has to members: an HttpListerContext class variable which includes request and response possibilities as well as a stream which stores the transferred data and a WebListener class variable to access the functions from the WebListener class. The only function implemented in the Worker class is ProcessRequest (). When instantiated in the WebListener class, the instantiates WebLister itself and the accepted context have to be passed to the constructor. In the ProcessRequest () functions the context is extracted from the stream and the data is logged by calling the WebListeners Write2Log () function. After successful reading from the stream an acknowledgement is sent back to the remote sender.

### 4.2.6 Camera Control

Camera control can be accessed when activating the camera tab in the form. It automatically starts capturing the camera images and displays the received images in a picture box. While the connection to the camera is being established, a bitmap that is displayed in the picture box notifies the user about the process. If no connection is available, the user is notified by a different bitmap that is loaded into the picture box. For camera access the camera IP and the URL must be specified and the necessary network credentials (username and password) must be created using the NetworkCredential class in the System.Net namespace.

The camera control part implements two main functions: ProcessImage () and GetImage () as well as two functions for creating the notification bitmaps mentioned above. The ProcessImage () is triggered using a new thread inside the main form. This thread is started when the camera tab is selected and stops when any other tab is activated. It captures images as long as a Boolean capture variable is set to true. To avoid over- head, the Sleep () function inside the System.Threading namespace should be called after capturing an image. An integer number can be passed to this function that indicates how many milliseconds the thread should sleep. Changing this value will also change the sampling rate of the camera, if a slower or faster sampling rate is required.

Due to the high sampling frequencies for live streams, memory overflow may occur. The .NET garbage collector works too slowly in this case, though it is possible to call it inside C# code. This is done as follows:

GC.Collect (); - Calling this function will remove all unused references and memory allocated in the image acquiring process. This function should be called after each image captured. The ProcessImage () function calls the GetImage () by passing an URL as string where to load the image. The GetImage () function creates a new WebRequest from the System.Net namespace using the URL provided from the ProcessImage () function. The image information is stored in the web request stream and can be extracted by calling the static FromStream () function inside the Image class. A code example is provided below:

*string URL = FileURL;*

*WebRequest request = WebRequest.Create(URL);*

*request.Proxy = null;*

*request.Credentials = new NetworkCredential(camUser, camPassword);*

*request.Timeout = 5000;*

*Image CamImage = Image.FromStream(request.GetResponse()*

*.GetResponseStream(), true, true);*

The two 'true' values passed to the function enable embedded colour management and image data validation. The extracted image is then displayed inside the form using the appropriate delegate inside the form. The functions for creating the notification bitmaps are used when starting the thread before the first image is acquired or when the WebRequest instance inside the Getimage () function returns a timeout for the request.

## 4.2.7 Database Design & Implementation

A datab a s e and its tables are usually designed using the Entity-Relationship-Attribute-Model, short ERA. The ERA uses parts of the real world that are then mapped to data models. There are three main parts to describe these mappings: entities, relationships and attributes. Entities for example could be persons or projects. In general, these entities are objects in the real world. Relationships are used to link different entities or to create context between entities. Attributes are information of interest about entities, for example the name and birthdate of a person. This model is used in further steps to create a database that fits the needs for the setup.

### 4.2.7.1 Conceptual Data Model

For designing the database model, Sybase PowerDesigner 15[7] is used. First, a CDM (Conceptual Data Model) has to be created as it can be seen in A. This type of model covers the structure and planning of a database. From a CDM either a PDM (Physical Data Model A) or a LDM (Logical Data Model A) can be derived.

The CDM consists of several tables, also called entities. These entities are connected via relationships. Possible relations are one to one (1:1), one to many (1:n) and many to many (m:n). These relations may result in restrictions or new tables that occur in the PDM or LDM. The following entities were created in the CDM:

- SensorTypes
- SensorNodes
- SensorMeasurements
- SensorData
- Person
- Rooms

---

- Cameras
- CameraImages

In this database prototype only important information for each entity was included. Additional information can be easily added using Sysbase PowerDesigner.

### 4.2.7.2 Entities

The 'SenorTypes' entity is used to specify the type of a sensor node when different types of Sensors are used, e.g. pulse sensors, accelerometers or temperature sensors. Each type is described by a maximum of 64 characters or numbers and has a unique integer number as an ID which also serves as the primary key for this table. This ID is also used to link the database entries in the 'SensorTypes' entity to the entries in the 'SensorNodes' entity.

The 'SensorNodes' entity is used to store all information about every sensor. Each sensor has a unique ID and a unique address. The ID is entered as an integer number and the IEEE address may contain a maximum of 64 characters or number. The ID and the IEEE address are used as the primary key for the 'SensorN- odes' entity. It is also possible to name each sensor. The name is stored in a field of maximum 64 characters.

The 'SensorMeasurements' entity groups several entries from the 'SensorData' entity to a single measurement. A grouping of several sensor data values is required due to the fact that there may be sensors with more than one value measured, e.g. an accelerometer measures acceleration in x, y and z directions when a Cartesian coordinate system is used. For each measurement, a timestamp is stored as well. As there may be several measurements at the same time, this timestamp is not unique for a single measurement and therefore may not be used as the primary key. Each measurement therefore has a unique ID and is stored as an integer. Additionally, a flag called 'risk' can be set to identity whether certain values of a measurement were above or below given thresholds.

The 'SensorData' entity stores all kinds of sensor values. Each sensor data has a unique ID that is an integer number. This number also serves as the primary key for the table. Each of the sensor data values is stored in a single field that is a float data type.
The 'Persons' entity stores information about people who carry a sensor, e.g. patients in a hospital. For the sake of convenience each person listed in the Persons entity has an ID that is an integer number and the persons first and last name. The ID serves as the primary key in this table and the names may not be longer than 64 characters. This entity could cover more detailed information like address, illnesses and so on.

The 'Rooms' entity is used for storing information about rooms. This is achieved by the use of an ID (integer number) and a name with a maximum of 64 characters. The ID is the primary key for the 'Rooms' entity. This entity is mainly used for linking sensors that are carried by persons to cameras that are in the same room. So, for example it is now possible to store a camera image to the 'CameraImages' entity when a certain sensor value is measured by knowing which camera is in the same room of the sensor.

The 'Cameras' entity is used to store information about all deployed cameras. Each camera has a unique integer ID that serves as the primary key for this table. This table also stores the name with a maximum of 64 characters and the camera IP with a maximum of 16 characters. This table also includes the URL for acquiring the camera images that is used by the web server application. This URL may not have more than 128 characters. This table may be extended with further camera commands if required.

The 'CameraImages' entity serves as storage for all camera images that need to be stored or archived. Each image has a unique integer ID that is also as the primary key for this table and a timestamp when the image was taken. This timestamp may not be unique as there are several cameras that may take a picture at the same time and therefore is not suitable for primary key purposes. The 'image data' field contains binary information of an image that was stored to the database using streams. A full image can be obtained my deserialising the binary information.

### 4.2.7.3 Relationships

In order to connect entities and have SQL queries to function correctly, relationships between database tables or entities have to be defined. As mentioned above there are three main relationships:

- One-to-one
- One-to-many
- Many-to-many

These relationships may be slightly modified by mandatory or dependent entities or minimum requirements to be zero or one. The first relationship is located between the 'SensorTypes' entity and the 'SensorNodes' entity. A type of a sensory may describe zero, one or more sensor nodes while a sensor node may only have one sensor type describing it and a sensor node may not exist without a sensor type describing it. This finding results in a one-to-many relationship from 'SensorTypes' to 'SensorNodes' having the 'SensorTypes' entity set to dependent for the 'SensorNodes' entity. The second relationship arises between the 'SensorNodes' entity and the 'SensorMeasurements' entity. A sensor node may have collected zero, one or many sensor measurements.

A sensor measurement may only have been taken by one certain sensor node and may not exist without a sensor node. This requires the 'SensorNodes' entity to be set dependent for any 'SensorMeasurements' entry. This relationship is a one-to-many relationship when seen from the 'SensorNodes' entity's view.

The next relationship occurs between the 'SensorMeasurementss' entity and the 'SensorData' entity. A sensor measurement may consist of at least one or more sensor data values. Sensor data may only be part of one certain sensor measurement and may not exist without a sensor measurement. This requires the 'SensorMeasurements' entity to be set dependent for any 'SensorData' entry. This relationship is a one-to-many relationship when seen from the 'SensorMeasurements' entity's view.

A further relationship is located between the 'SensorNodes' entity and the 'Persons' entity. A sensor may be attached to zero or one person(s) while a person may carry zero, one or more sensors, so this is a one-to-many relationship when viewed from the 'Persons' entity to the 'SensorNodes' entity.

There is a further relationship between the 'Persons' entity and the Rooms' entity. A person may live in one or many rooms, whereas a room may have zero, one or more persons in it. This results in a many-to-many relationship. These kinds of relationships require an additional entity that is created in the LDM. This new entity will store all additional information about the connected persons and rooms.

The next relationship is between the 'Rooms' entity and the 'Cameras' entity. Each room may have zero, one or more cameras mounted whereas a camera can only be mounted in zero or one room(s). This finding leads to a one-to-many relationship from the 'Rooms' entity's view.

The last relationship exists between the Cameras entity and the 'CameraImages' entity. Cameras may take zero, one or many images. An image may only refer to one camera and a camera image may not exist without a camera that has taken it. So the 'Camera' entity has to be dependent on the 'CameraImages' entity. So this is a one-to-many relationship seen from the 'Cameras' entity to the 'CameraImages' entity. All other entities do not have any relationship with each other.

### 4.2.7.4 Logical Data Model & Physical Data Model

The physical and logical data model can be easily derived from the conceptual data model using Sybase PowerDesigner. To start the conversion the conversion tool must be started via 'Tools' - 'Generate Physical Data Model' or 'Tools' - 'Generate Logical Data Model' inside the PowerDesigner application.

The logical data model now contains the database entities, as they will be stored inside the database server, see appendix A. It can be seen that the relationship between the 'Rooms' entity and 'Persons' entity resulted in a complete new entity that is connected to the 'Rooms' and 'Persons' entities. Furthermore, all dependent relationships lead to an extension of the respective entity. For example the 'SensorData' entity now contains primary key fields from the 'SensorMeasurements' entity that serve as foreign keys as well as primary key in this table. The 'SensorData' even contains primary key fields from all sensor entities due to the dependency property set in these entities. The nondependent relationships only turn into foreign key in the respective entity, for example the ID of the 'Rooms' entity becomes a foreign key in the Cameras' entity. These foreign keys are then used to join tables for specific SQL queries, so that only connected table entries are selected.

The physical data model derived from the CDM is fitted to a certain data base architecture, e.g. different versions of the Microsoft SQL Server, Oracle database server or MySQL server. It uses data types available from the selected database architecture. Due to experience, the Microsoft SQL Server is chosen to be the storage device of the prototype. The model described before may be easily imported to the Microsoft SQL Server 2008 Express using the Sybase PowerDesinger tool. This tool can be launched via 'Database'- 'Generate Database' inside the PowerDesigner application. This tool generates a full SQL script file that is then executed on the Microsoft SQL 2008 Express database server. Now it is possible to fill the database tables with information about persons, rooms, sensors and cameras using Microsoft's SQL Server Management Studio that is included in the installation package.

Microsoft offers a free lightweight version of the Microsoft SQL Server 2008 known as Microsoft SQL Server 2008 Express. This express server fits all prototype needs. The RC2 of the express server is used. The SQL server can be accessed from C# and other programming languages using the SQL server native client driver.

### 4.2.8 SIP Server

3CX phone system is a free Session Initiation Protocol (SIP) server[8]. It can be configured via a HTTP interface for remote configuration or a management tool. The interface is easy to handle and new users can be created without great knowledge of SIP and SIP servers. The server handles all SIP communication for created users, e.g. cameras and cell phones. A typical setup scenario[9] of a SIP server can be seen in figure 4-6.

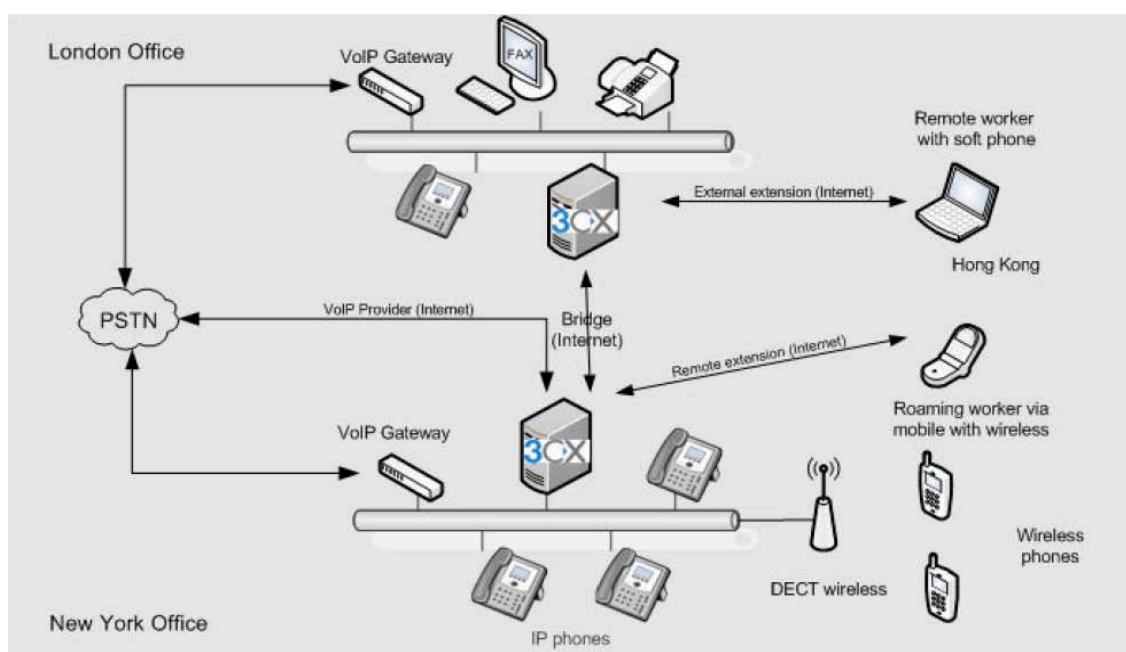

**Figure 4-6: Typical SIP Server Setup**

From this setup, the idea was derived to use a mobile phone with wireless LAN capability. Due to the available material for the prototype, an iPhone was used. Regarding SIP software on the phone, two programs were tested: NetDial Sip Phone[10] and PocketBone[11]. The first one was more stable than the second one, but due to the fact that only PocketBone supports video calls over SIP, it was chosen for the prototype despite its instability.

---

[8] http://www.3cx.com/ip-pbx/index.html
[9] http://www.3cx.com/phone-system/3CXPhoneSystem brochure.pdf
[10] http://itunes.apple.com/us/app/netdial-sip-phone/id378721185?
[11] http://itunes.apple.com/us/app/pocketbone/id349891336?

### 4.2.9 Network

The whole network uses TCP/IP communication except the wireless part between sensors and base stations. It is possible to use the setup in an Ethernet LAN or separate parts and connect them over the Internet. For example the web server and the database can be deployed and accessed via the web server application. Beware that firewalls may block connection establishments if they are configured incorrectly! The sensor information may then be sent over the Internet by the base station using HTTP. Base stations and sensors always have to be connected locally.

### 4.2.10 Programming

In order to develop the software for the prototype as well as for the energy consumption simulations, the predefined package was used that comes along with the SPOTs. It consists of Solarium as a testing environment, NetBeans 6.9.1 as IDE and all necessary drivers for the USB connection. The developing hardware was a 2006 MacBook Pro running Mac OS 10.6.5. Before starting, one has to choose the SDK version for flashing and programming the SPOTs. In this thesis, the v6.0 Yellow Beta2 SDK was used. The first step before the actual programming and deployment can start is to flash the SPOTs and also the USB transceiver unit to the SDK used for development. Connecting the devices via USB and download the SDK onto them by using the SPOT Manager Tool that comes along with Solarium can do this.

The software for the prototype is based on a re-written combination of the Telemetry-Demo, SendDataDemo and HTTPDemo included in the SDK. It consists of three Java classes called AccelMonitor.java, TelemetryMain.java and a helper class called PeriodicTask.java. The last mentioned class is taken as a whole out of the TelemetryDemo and provides a framework for a task in order to take samples in a regular interval. It makes use of the AT91 TC timer counter of the SPOTs with reliable results as long as the interval is smaller than two seconds.

The AccelMonitor.java class is responsible for gathering the raw data out of the LIS3L02AQ Accelerometer on board the SPOTs. It implements the former mentioned PeriodicTask.java class and therefore has a doTask () method that is triggered periodically to obtain the accelerometer values. Out of these values, the total square of all three axes is computed and checked by the method doCheckValue () in order to verify if the measured values exceed the threshold values that indicate a fall of a person. This threshold was obtained empirically beforehand via test series using the unmodified TelemetryDemo and its graphical desktop counterpart. For an example graph see figure 4-7. If the threshold is violated, the method doSendHTTP () establishes a HTTP connection to the web server in order to transmit the alarm signal.

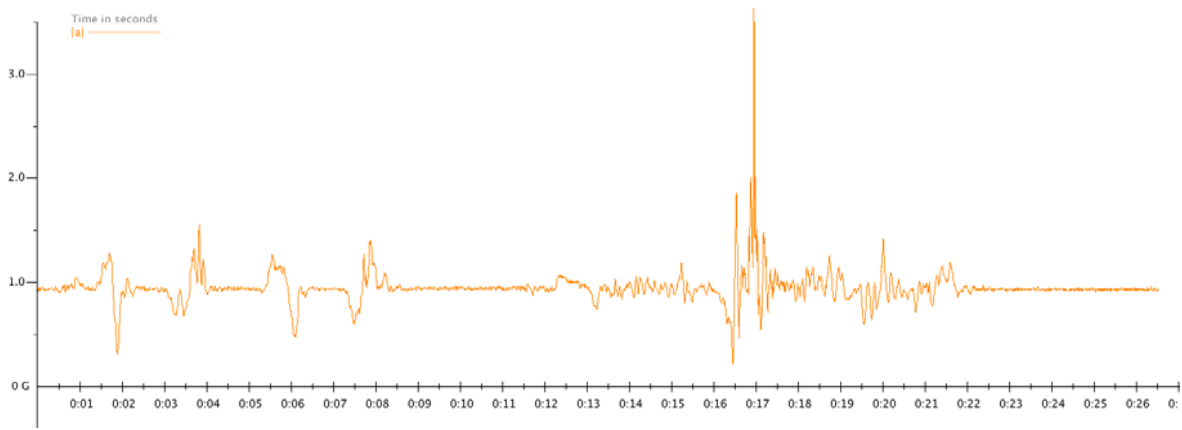

**Figure 4-7: Example Graph of the Used Telemetry Demo**

The third class named TelemetryMain.java is extending the MIDlet class from the Java ME in order to be run-able on the SPOT's VM. Once the SPOT is turned on, TelemetryMain.java is automatically loaded after booting.

Unless the SPOT is turned off or the Java program is stopped via Solarium, it runs on constantly. In the setup for the simulations regarding the energy consumption of the SPOTs, the programming structure was altered in comparison to the prototype mentioned before. One change was a new implementation of the periodic sampling without the use of the PeriodicTask.java in order to reduce the footprint of the application and use as much self-written code as possible.

This time the code package consists of two classes, DropMonitor.java and SensorSampler.java. The first one includes access to the accelerometer and provides a method called getValues () to read out the raw values together with the current battery capacity in mAh. Concerning the battery capacity, there arose some difficulties to obtain these values. It seems that in every SDK the way of accessing the battery has changed. One would assume that like any other class an instantiation would look like this:

*IBattery myBattery = new Battery ();*

However when it comes to the SPOTs it is necessary to do this via an instance of the board in the SPOT. From this instance the PowerController object has to be gained and from this in return a Battery object. Interestingly in the project for the prototype it was possible to do this in one rush:

*private IBattery = Spot.getInstance().getPowerController().getBattery();*

In the project of the simulations it had to be done in two steps. Until now, the authors were not able to figure out why this misbehaviour occurs. The handling of the battery seems to be a general problem, as there are a lot of discussions going on in the SunSpot forum as well.[12][13]

---

[12] https://www.sunspotworld.com/forums/viewtopic.php?f=37&t=3551

[13] https://www.sunspotworld.com/forums/viewtopic.php?f=37&t=1428

```
private IPowerController power = Spot.getInstance().getPowerController();
private IBattery battery = power.getBattery();
```

The SensorSampler.java class contains an instance of the above described DropMonitor.java class. It also holds the three different intelligence levels called Tiers for the according simulations described in a section later on. The Tier implementations al- low the simulation of normal IEEE 802.15.4 radio transmissions as well as tests with additional layers like TCP used for HTTP.

## 5 Experiments and Results

In order to evaluate a sensor node's energy consumption regarding its given intelligence, a tier-based concept is introduced. Figure 5-1 shows that a low intelligent sensor node requires a certain amount of energy. Energy consumption is decreasing while sensor nodes intelligence is increased.

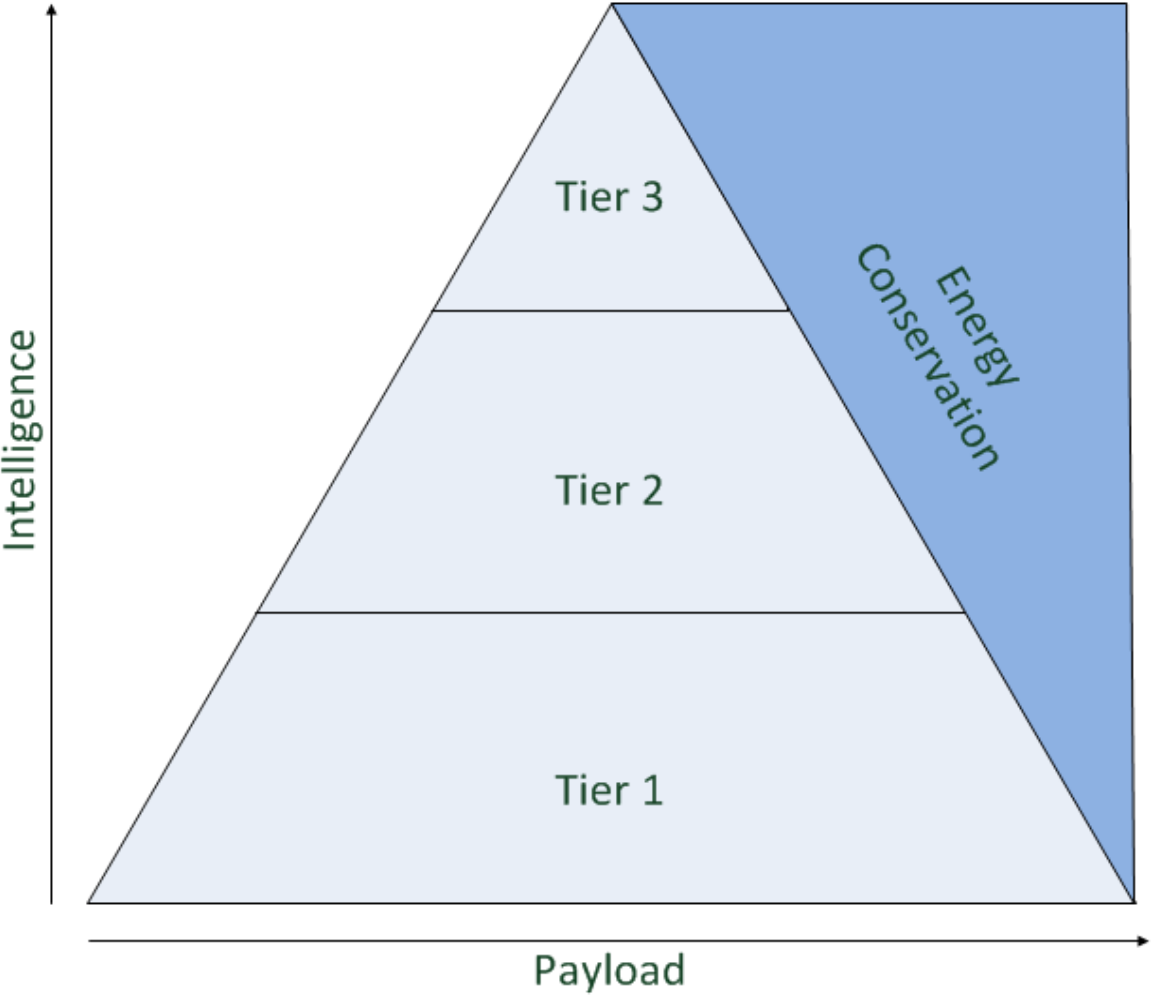

**Figure 5-1: Intelligence of the Sensor vs. Sent Data**

For example, if sensor measurements are processed and evaluated on a remote workstation, each sensor data value must be transmitted from the sensor node to its base station, which results in a high payload.

Due to the fact that energy consumption is higher while transmitting than when processing data on the CPU or reading sensor values, it is obvious to reduce the amount of transmissions and rather increase processing time on the sensor. This is done by implementing logic, in this case intelligence, on the sensor itself. The sensor is then responsible for pattern recognition and decision making itself. Data transfer only occurs if required, e.g. data values match a certain pattern.

As mentioned before, a tier-based intelligence system is introduced. A higher tier of intelligence refers to a more complex version of the nodes intelligence implementation. The energy consumption of each tier is then measured and evaluated.

## 5.1 Sensor Node Intelligence Concept

A three-tier intelligence system is chosen to fit all necessary requirements for energy consumption measurement. Intelligence implementation is kept simple in each tier since measuring and evaluating a sensor nodes energy consumption behaviour is prior in order to making a node as intelligent as possible. The approaches described in detail in the next three sections were chosen as the three intelligent tiers. Figure 5-2 displays a graph indicating which sensor values are transmitted by each tier.

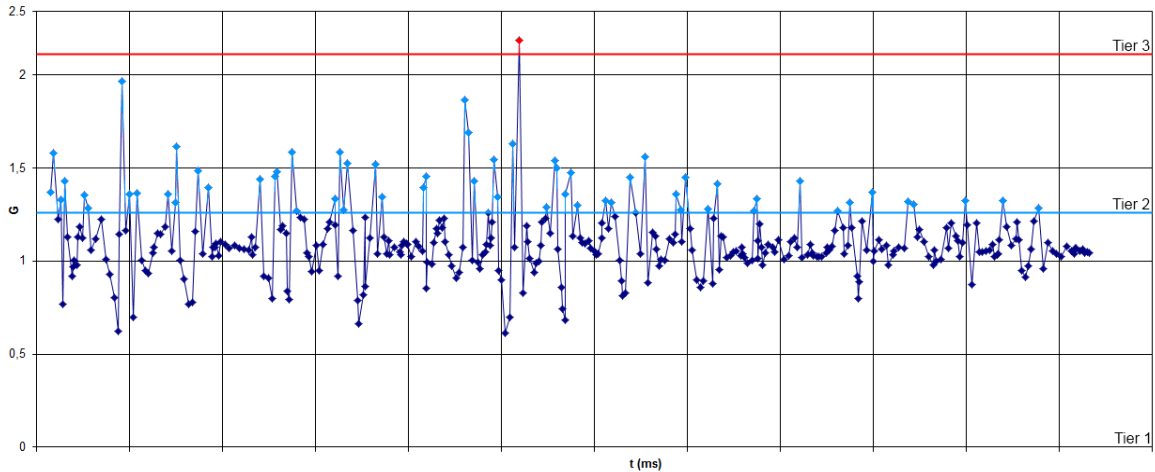

**Figure 5-2: Data Selection According to Each Tier Level**

➢ **Tier 1 - The non-intelligent sensor node** – The Tier 1 sensor node was implemented to send each measurement immediately after it gets measured. This can be seen in figure 5-2. No processing intelligence was implemented and thus the sensor node itself does not make a decision itself. As there are no decisions made by the sensor node, this task will be carried out by a remote workstation which has intelligence implemented. It is expected to be the most energy consuming method because the radio communication will be active permanently. However, if communication interruptions occur, these interruptions can be used as signals to determine different situations. For example, the person carrying a sensor walked outside the observed area, the remaining sensor battery could be too low or the sensor device had been deactivated unintentionally. Therefore, it can be pointed out that in either case something has gone wrong and the observed person requires assistance.

➢ **Tier 2 - The semi-intelligent sensor node** – The Tier 2 sensor node is able to decide on its own whether it is actually necessary to transmit data to the base station or not. This decision is based on a person's movement that is measured by the sensor node. For example, if the person starts moving, like standing up, walking or sitting down data transmission to the base station is initiated. If the movement stops, for example when the person is sitting or resting, the sensor will stop its transmission. In this case, the risk of falling and other dangerous situation is minimal when compared to movement situations. A threshold defines whenever data transfer is necessary. This can be seen in figure 5-2. This semi- intelligent tier is intended to save battery energy due to the reduced amount of payload. However, for pattern recognition and decision making a remote work- station is required. The energy consumption in this implementation is expected to decrease as of the interrupted transmission during periods of time in which the person carrying a sensor is not moving. Observation of devices functionality like remaining battery or the person's location is almost impossible. One possibility to overcome this drawback is to include a periodic heartbeat signal that would lead to a higher payload and therefore higher energy consumption. This was not implemented in the prototype in order not to distort measurements.

➤ **Tier 3 - The intelligent sensor node** – The implementation of the intelligent sensor node fulfils simple pattern recognition itself. A single alarm message is transmitted each time a person's fall is detected instead of transmitting several data values in a certain time period. Therefore, a higher threshold than in the previous tier is defined. This is a very basic solution that can be easily extended or modified if required. The remote workstation is directly informed about the sensor nodes decision, e.g. that the person fell. The measured data values are not transmitted as they are in the two previous tiers. Therefore, the transmission only occurs in emergency cases. However, the node itself needs to be capable of recognizing data patterns. In some cases this leads to the need of a powerful processor on the sensor node, depending on data complexity and feature extraction algorithm. Due to this fact, the pattern recognition intelligence of this tier is kept simple. This kind of implementation is expected to have the lowest energy consumption. Otherwise, the sensor device is only observing and processing data. To observe the sensors functionality, a periodic heartbeat signal would be needed (see previous section).

## 5.2 Discussion of Results

In order to ensure a deterministic behaviour, the simulations were done in three runs for each tier. Tables 5.1 to 5.3 illustrate the data extracted from the simulation runs. All tiers use the IEEE 802.15.4 standard regarding radio communication. As performance evaluation criteria the energy drain of the sensor node for each run was measured in coulomb per minute. As it turned out, the simulation outcomes regarding the energy consumption within the single tiers were all within a certain range except the second run of tier 1, which could be traced back to errors in the sensor readings.

By comparing the outcome of tier 1 to the outcome of tier 2, it can be seen that tier 2 results in 12.7% less energy consumption than tier 1. This result can be traced back to the reduction of the transmitted sample data. This has been expected due to the decrease of sent data in average from 746.8 to 37.2 transmissions per minute. Hence, 95.1% less data was transmitted - from 173,515 transmissions to 6,532 in average.

**Table 5.1: Battery Consumption Overall Table Tier 1**

|            | Total      | per minute | data   | send rate | samples |
|------------|-----------|------------|--------|-----------|---------|
| **Tier1-Run1** | 2717.13 Ws | 0.1907W   | 176902 | 80.57ms   | 176902  |
| **Tier1-Run2** | 2328.6 Ws  | 0.1711W   | 168576 | 80.75ms   | 168576  |
| **Tier1-Run3** | 2655.45 Ws | 0,1885W   | 175066 | 80.45ms   | 175066  |
| **max**    | 2717.13 Ws | 0.1907W   | 176902 | 80.75ms   | 176902  |
| **min**    | 2328.6 Ws  | 0.1711W   | 168576 | 80.45ms   | 168576  |
| **average**| 2567.02 Ws | 0.1834W   | 173515 | 80.59ms   | 173515  |

**Table 5.2: Battery Consumption Overall Table Tier 2**

|  | Total | per minute | data | send rate | samples |
|---|---|---|---|---|---|
| **Tier2-Run1** | 1628.37Ws | 0.1582W | 5341 | 80.13ms | 202729 |
| **Tier2-Run2** | 1698.15Ws | 0.1594W | 10641 | 80.45ms | 206707 |
| **Tier2-Run3** | 1741.85Ws | 0.1627W | 3615 | 79.56ms | 211987 |
| **max** | 1741.85Ws | 0.1627W | 10641 | 80.45ms | 211987 |
| **min** | 1628.37Ws | 0.1582W | 3615 | 80.13ms | 202729 |
| **average** | 1689.49Ws | 0.1601W | 6532 | 80.05ms | 207141 |

The number of transmissions in tier 3 is reduced to 8 in average from tier 2 to tier 3, which results in a reduction of 99.8%. However, in numbers of transmissions it is only a reduction of about 6,524 that is compared to the reduction from tier 1 to tier 2 only of 3.9%. It leads to a slight increase, which means a reduction in energy consumption of 11.4% in comparison to tier 1.

Furthermore the Watt-second per sample are 14.7942 mWs for tier 1 and 8.1662 mWs for tier 2 (44.76% reduction compared with tier 1) and 8.1326 mWs for tier 3 (44.94% reduction compared with tier 1).

Overall, the resulting energy savings comparing tier 3 and tier 2 did not reach the authors' expectations, though the savings in data transmissions are minor compared to the 95.1% data transmission reduction from tier 1 to tier 2.

The deviant results in tier 3 regarding the expected energy consumption are interesting and important issues. Some margin of the gained divergence can be interconnected to the experimental technology platform SunSPOT. However, as it was demonstrated in research before, a significant energy drain can be observed during idle and overhearing periods.

**Table 5.3: Battery Consumption Overall Table Tier 3**

|  | energy consumption | | data | samples |
|---|---|---|---|---|
| **Tier3-Run1** | 2314.35Ws | 0.1604W | 9 | 288411 |
| **Tier3-Run2** | 2333.41Ws | 0.1610W | 11 | 289871 |
| **Tier3-Run3** | 2397.86Ws | 0.1665W | 4 | 288072 |
| **max** | 2397.86Ws | 0.1665W | 11 | 289871 |
| **min** | 2314.35Ws | 0.1604W | 4 | 288072 |
| **average** | 2348.58Ws | 0.1627W | 8 | 288785 |

This drain can be as high as transmitting or receiving data [9, 35, 34]. Therefore, it can be stated that it is not possible to reduce the overall energy consumption related to the radio device below a certain value. Though, a total shutdown or a deep sleep cycle of the radio part would be a suitable compensation in this demonstration setup, due to the fact, that the sensor node is only sending data and receives none.

The authors like to remark, that in this thesis the hardware was used as it was without any physical manipulation. Therefore, only the resources available in the SunSPOT API were used in order to do the measurements. After evaluating the results according to the measurements done, the authors came to the conclusion that a more appropriate and accurate way would be in performing the measurements physically by using external instruments which is subject to future work.

## 5.3   Discussion of an Alternative Approach

Due to the fact that the amount of energy that could be saved was smaller than expected, an alternative approach to the simple threshold concept may be considered. Therefore, the authors of this thesis suggest the concept of neural networks, which could be implemented on the sensor in exchange for the threshold software concept.

The idea of artificial neural networks is based on biological neural networks [2]. The term neural network in this text refers to the artificial version of a neural network. In order to achieve a similar structure and functionality of its biological equivalent, a neural network consists of several artificial neurons that are interconnected. Each connection between two nodes in a network is weighted individually. Therefore, neural networks are able to 'learn' input patterns through different algorithms as explained in [4] and [14] and perform input classification. These learning algorithms can either be supervised or non-supervised. Supervised networks require a learning cycle before meaningful classification is achieved, whereas non-supervised perform their learning on the fly. The most basic form of a neural network is a simple perceptron that operates as a linear two-class classifier as described in [33]. It uses a single binary output ('Heaviside' function) for classification.

An input is defined as a vector xi (i=0...N-1) where N is the number of inputs to the network. Input values are weighted by corresponding weights of a vector wi (i=0...N-1) and summed up. A bias b may be added. The output y is therefore defined as follows:

$$y = \begin{cases} 1 \rightarrow \sum_{i=0}^{N-1} \omega_i * x_i + b \geq 0 \\ 0 \rightarrow \sum_{i=0}^{N-1} \omega_i * x_i + b \leq 0 \end{cases}$$

$$(5.1)$$

A perceptron is trained using the perceptron-l e a r n i n g algorithm. New values for weights and bias are calculated as follows where t is defined as the target value (0 or 1).

$$\omega_i^{new} = \omega_i^{old} + (t - y) * x_i$$

$$(5.2)$$

$$b^{new} = b^{old} + (t - y)$$

$$(5.3)$$

An advanced form of a perceptron is the so-called ADALINE (Adaptive Linear Element) developed by Widrow and Hoff (Widrow & Hoff, 1960), which uses the least-mean-square (LMS) learning algorithm and a sign (+1 and -1) output function. This algorithm changes the weight values of the vector w in order to minimize the squared error. The error function Q relative to a weight $w_i$ is shown in figure 5-3.

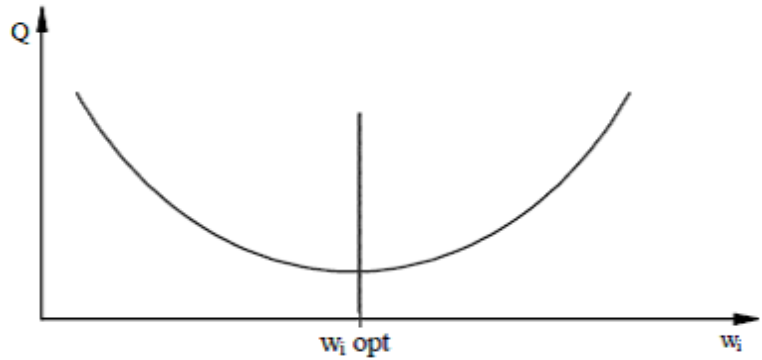

**Figure 5-3: Graph of Error Function Q**

The error function Q is defined as follows:

$$Q = \overline{e^2} = \frac{1}{M} * \sum_{m=1}^{m-1} [\![(y]\!]_m - t_m)^2 \qquad (5.4)$$

The value for M is defined as the number of different input vectors. Usually the weights are adjusted after each step in a learning cycle where M = 1. In some cases, M is chosen larger than 1, which is called 'batch' learning. Q is derived by $w_i$ using the chain rule in order to move the weight $w_i$ towards its optimal value depending on the gradient of Q. A scaling factor η is introduced to determine the weights'adjustment. This factor is often referred to as the learning rate. This leads to the following overall result (assuming M = 1):

$$\omega_i^{new} = \omega_i^{old} + \eta * (t - y) * x_i \qquad (5.5)$$
$$b^{new} = b^{old} + \eta * (t - y) \qquad (5.6)$$

Perceptron and ADALINE have the shared restriction of classifying only linear separable problems. In order to extend these networks to non-linear classification the output of either perceptron or ADALINE are changed to sigmoidal (non-linear) functions with proper derivatives and combined to a complex network of several elements. The simplest form of these networks is called back-propagation network that is a multilayer perceptron and described in [48].

These networks are based on a generalized version of the LMS algorithm (gradient decent algorithm) which allows non-linear transfer functions for each unit as long as they are derivable. A network usually consists of three layers: input, hidden and output layer. Each input node is connected to each hidden node as well as each hidden node is connected to each output node. An example for a back propagation network can be seen in figure 5-4.

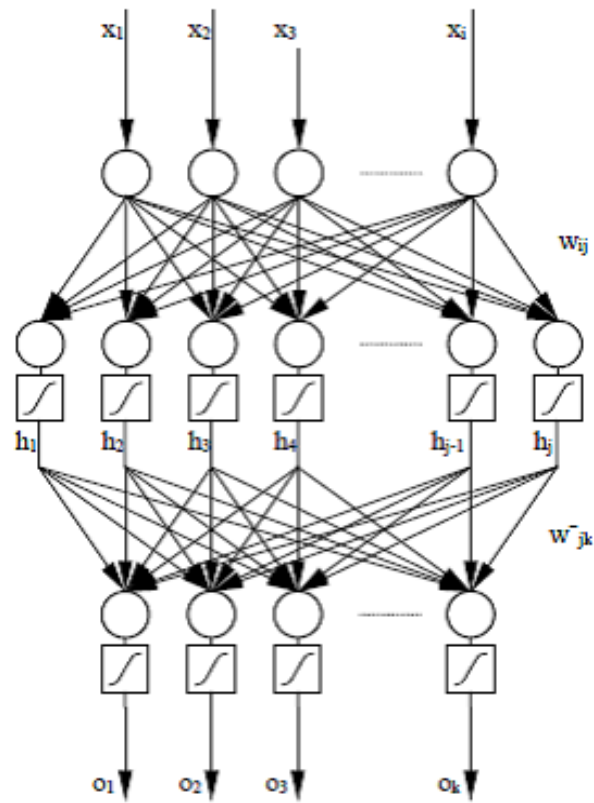

**Figure 5-4: Example of a Back Propagation Network**

The error in back propagation networks propagate backwards from the output nodes to the inner nodes as described in [47] and weights that minimize the error are adjusted. These networks and the knowledge how they operate can be used to detect movement patterns in sensor node measurements.

Each input vector covers a certain amount of samples collected by the sensor node. The number of input nodes must match the number of samples in the input vector. It is important that all possible patterns can be represented within the samples provided as input for the network! Each output node matches a decision towards a specific pattern that is depending on the use of the sensor node.

In the case of movement detection examples for these patterns would be: walking, running and falling. Decisions are transmitted to the sensor's base station either as strings or for simplicity as single integer values to reduce the transmission amount. Training data must be gathered before implementing a back propagation network on the sensor node. A dedicated computer can be used to calculate all necessary weights for the network, which are transferred to the sensor node after calculations are completed to avoid training cycles on a sensor node itself and to preserve energy for measurements.

There are several frameworks available for implementing neural networks on different platforms such as C# or Java. A Java framework is provided by Oracle.

# 6 Conclusion

Around the globe the number of older people in relation to the rest is constantly growing. As a result, medical and care facilities cannot handle the growing number of patients. Therefore, elderly in-home assistance gets more attention and importance. Due to issues regarding memory, physical strength and reduced self-assessment, old people face a lot of challenges in accomplishing their activities of daily living.

This thesis was meant to address the above-mentioned issues in two ways. First, a prototype was developed that provided functions for surveillance of elderly people in ambulant home-care, retirement homes or hospitals. The main aspect of the prototype was to implement an alert system in case a patient falls. If such an accident occurs, the nursing personal or relatives should be able to contact the patient and check remotely whether she is all right or medical treatment is necessary. Using a wireless sensor that has embedded an accelerometer in order to detect the fall and in case sends an alert signal to a web server did this.

The server then triggers a 360-degree sight camera with an intercom. The camera initializes a SIP call in order to get help and clarify the situation. Also, normal room surveillance is possible by using the web server and does a motion habit analysis of the gathered patient data in a graphical way. Long-term analysis is also possible due to a database accessible by the web server.

The second part of the thesis work focused on the energy consumption issue of the used wireless sensor node. It could be shown that it is possible to reduce the energy consumption in certain application scenarios by increasing the sensor node's intelligence and therefore reduce the amount of data and the used energy for acquiring and transmitting it. The conducted simulation has shown that, with an increase of intelligence in the sensor node, a data reduction of over 95% can be achieved which is a dramatically reduction. Also, a reduction of energy consumption could be increased by 12%.

While this is below the expectations the authors had, it still is a noticeable decrease. It clearly indicates that increasing intelligence in sensor nodes is only one possible step to solve the energy consumption issue and not the ultimate solution. Additional concepts like the described SmartReflex Technology approach by Texas Instruments or energy scavenging concepts should be considered and integrated into the system concept.

# 7 Future Work

Some interesting points for further development regarding the prototype as well as the optimization of the sensor's energy consumption exist. Possible add-ons for the prototype would be the implementation of supporting additional sensors for other purposes like blood pressure or pulse and their online analysis with help of the web server. Also, a multi-patient control centre would be attractive for use in hospitals or retirement houses.

Regarding the energy savings of the sensor, one could conduct further research in the difference of using other frequency bands as 2.4 GHz as well as other radio technologies like Bluetooth. Another interesting scenario would be the use of multiple sensor nodes with a different offset regarding the used sampling time. The resulting energy savings compared to the higher costs for the additional sensors would be interesting. Also, the actual implementation of the neural network approach as a comparison remains future work.

In addition to that, the topic of energy harvesting in order to power the sensors battery while on duty is an important aspect. Due to the fact that the sensor in this thesis is worn at the belt in the middle of the body, it is normally covered by cloth and therefore solar panels or similar are not suitable. One could think of charging via a magnetic field or by mechanical energy like it is used for hand watches in order to draw them.

As already mentioned in the thesis before, the measurement of the energy drain could also be improved by the use of external instruments like an Ampère meter in order to get more accurate data.

# 8 References

[1] Alemdar, Hande and Ersoy, Cem: Wireless sensor networks for healthcare: A survey. Computer Networks, 54(15):2688 – 2710, 2010, ISSN 1389-1286. http://www.sciencedirect.com/science/article/pii/S1389128610001398.

[2] Bar-Yam, Y.: Dynamics of Complex Systems. An online version is available at http://necsi.edu/publications/dcs/Bar-YamChap2.pdf (23.05.2011), June 2003.

[3] Best, R. and Valence, G. de: Design and construction: building in value. Elsevier, MA / USA, 2002.

[4] Bishop, C.M.: Neural Networks for Pattern Recognition. Oxford University Press, Oxford, 1996

[5] Capehart, B.L. and Capehart, L.C.: Web based enterprise energy and building automatin systems. The Fairmont Press, GA / USA, 2007.

[6] Carlson, B. and Giolma, B.: Smartreflex power and performance management technologies: reduced power consumption, optimized performance. An online version is available at http://focus.ti.com/pdfs/wtbu/smartreflexwhitepaper.pdf (02.05.2011).

[7] Clements-Croome, D.: Intelligent buildings: design, management and operation. Thomas Telford Publishing, London, 2004.

[8] Dini, Gianluca and Tiloca, Marco: Considerations on security in zigbee networks. In Sensor Networks, Ubiquitous, and Trustworthy Computing (SUTC), 2010 IEEE International Conference on, pages 58 –65, June 2010.

[9] Dong, Q.: Maximizing system lifetime in wireless sensor networks. In Information Processing in Sensor Networks, 2005. IPSN 2005. Fourth International Symposium on, pages 13 – 19, April 2005.

[10] Echelon: Introduction to the LONWORKS Platform, rev.2. An online version is available at http://www.echelon.com/support/documentation/manuals/ general/078-0183-01B_Intro_to_LonWorks_Rev_2.pdf (02.03.2011).

[11] Echelon: Oslo Street Lighting System Slashes Energy Use with LONWORKS Network. An online version is available at http://www.echelon.com/solutions/ unique/appstories/oslo.pdf (02.03.2011), 2008.

[12] Farahani, S.: Designing Zigbee Networks and Transceivers: The Complete Guide for RF/Wireless Engineers. Butterworth Heinemann, MO / USA, 2008.

[13] Gomez, C. and Paradells, J.: Wireless home automation networks: A survey of architectures and technologies. Communications Magazine, IEEE, 48(6):92 –101, 2010, ISSN 0163-6804.

[14] Haykin, S.S.: Neural networks: a comprehensive foundation. Prentice Hall, Upper Saddle River / NJ, 1999, ISBN 9780132733502.

[15] Hypp¨onen, H. et al.: eHealth strategy and RTD progress in Sweden. An online version is available at http://www.ehealth-era.org/database/documents/ERA_Reports/eHealthERA_CountryReport_SWEDEN_final%2017-09-07.pdf (16.04.2011), Sep. 2007.

[16] Jansson, M.: Homecare and Technology: Old Dreams - New Means? Licentiate Thesis - Lule University of Technology, 2005(27), May 2005, ISSN 1402-1757.

[17] java.net: Squawk. https://squawk.dev.java.net/ (last access:02.12.2010), 2010.

[18] K., Kinsella and Velkoff, V.A.: An Aging World: 2001. An online version is available at http://www.census.gov/prod/2001pubs/p95-01-1.pdf (16.04.2011), Nov. 2001.

[19] Knight, M.: Wireless security - how safe is z-wave? Computing Control Engineering Journal, 17(6):18 –23, 2006, ISSN 0956-3385.

[20] Lauterbach, C., Strasser, M., Jung, S., and Weber, W.: Smart clothes selfpowered by body heat. An online version is available at http://www.future-shape.de/publications_lauterbach/BodyHeatAvantex2002.pdf (16.05.2011), 2002.

[21] Leijdekkers, P., Gay, V., and Lawrence, E.: Smart homecare system for health tele-monitoring. In Digital Society, 2007. ICDS '07. First International Conference on the, page 3, jan. 2007.

[22] Leonov, V., Fiorini, P., Sedky, S., Torfs, T., and Van Hoof, C.: Thermoelectric mems generators as a power supply for a body area network. In Solid-State Sensors, Actuators and Microsystems, 2005. Digest of Technical Papers. TRANSDUCERS '05. The 13th International Conference on, volume 1, pages 291 – 294 Vol. 1, june 2005.

[23] Miller, G. and Spoolman, S.: Living in the Environment (Centage Advantage Books). Brooks/Cole Publishing Company, Pacific Grove / CA, 2011.

[24] Mini, Raquel A. F. and Loureiro, Antonio A. F.: Energy in wireless sensor net- works. In Garbinato, Benot, Miranda, Hugo, and Rodrigues, Lus (editors): Mid- dleware for Network Eccentric and Mobile Applications, pages 3–24. Springer Berlin Heidelberg, 2009, ISBN 978-3-540-89707-1.

[25] Mobotix: Hemispheric Q24. http://www.mobotix.com/ger_DE/Produkte/Kameras/Hemispheric-Q24 (last access: 02.12.2010), 2010.

[26] Mondal, N.I., Zaman, S.U., Al Masud, A., and Alam, J.: Comparisons of maximum system lifetime in diverse scenarios for body sensor networks. In Computer and Information Technology, 2008. ICCIT 2008. 11th International Conference on, pages 73 –78, dec. 2008.

[27] Olve, N. and Vimarlund, V.: Elderly healthcare, collaboration and ict - enabling the benefits of an enabling technology. VINNOVA Report VR - Swedish governmental Agency for Innovation Systems / Verket fr Innovatonssystem, 2006(05), June 2006, ISSN 1650-3104.

[28] Pei, Zhongmin, Deng, Zhidong, Yang, Bo, and Cheng, Xiaoliang: Application- oriented wireless sensor network communication protocols and hardware platforms: A survey. In Industrial Technology, 2008. ICIT 2008. IEEE International Con- ference on, pages 1 –6, 2008.

[29] Prial, F.J.: WIRING BUILDINGS FOR INTELLIGENCE. An online version is available at http://www.nytimes.com/1984/05/13/realestate/wiring-buildings-for-intelligence.html (4.03.2011), 1984.

[30] Purwar, A., Jeong, Do Do, and Chung, Wan Young: Activity monitoring from real-time triaxial accelerometer data using sensor network. In Control, Automation and Systems, 2007. ICCAS '07. International Conference on, pages 2402 –2406, oct. 2007.

[31] Raghunathan, V., Schurgers, C., Park, Sung, and Srivastava, M.B.: Energy-aware wireless microsensor networks. Signal Processing Magazine, IEEE, 19(2):40 –50, Mar. 2002.

[32] Renaud, M., Sterken, T., Schmitz, A., Fiorini, P., Van Hoof, C., and Puers, R.: Piezoelectric harvesters and mems technology: Fabrication, modeling and measurements. In Solid-State Sensors, Actuators and Microsystems Conference, 2007. TRANSDUCERS 2007. International, pages 891 –894, june 2007.

[33] Rosenblatt, F.: The perceptron: A perceiving and recognizing automaton. Report 85-460-1, Project PARA, Cornell Aeronautical Laboratory, Ithaca, New York,1957.

[34] Singh, S. and Raghavendra, C. S.: Pamas power aware multi access protocol with signalling for ad hoc networks. SIGCOMM Comput. Commun. Rev., 28:5–26, July 1998, ISSN 0146-4833.

[35] Stemm, M., Katz, R.H., and Y.H., Katz: Measuring and reducing energy consumption of network interfaces in hand-held devices. In IEICE Transactions on Communications, volume E80-B, page 1125 1131, 1997.

[36] Sun, T. et al.: Evaluating Mobility Support in ZigBee Networks. In Kuo, T W., Sha, E., and Guo, M. (editors): Embedded and Ubiquitous Computing: International Conference, EUC 2007(17-20 Dec.), Taipei, Taiwan, Lecture Notes in Computer Science, pages 87–100. Springer Berlin-Heidelberg, 2010.

[37] Sun Microsystems Laboratories/Oracle: The Squawk System - Preliminary Draft 2.1. https://squawk.dev.java.net/source/browse/*checkout*/squawk/trunk/doc/TheSquawkSystem-Sep02.pdf (last access:02.12.2010),2002.

[38] Sun Microsystems Laboratories / Oracle: Sun SPOT World.

http://www.sunspotworld.com/docs/Red/SolariumUsersGuide.pdf (last access:02.12.2010), 2010.

[39] Sun Microsystems Laboratories / Oracle: Sun SPOT World - Getting Started. http://www.sunspotworld.com/GettingStarted/ (last access: 02.12.2010), 2010.

[40] Ting-pat, A. and Chan, W. L.: Intelligent building systems. Kluwer Academic Publishers, MA / USA, 1999.

[41] Vass, Dorottya, Vincze, Zoltn, Vida, Rolland, and Vidcs, Attila: Energy efficiency in wireless sensor networks using mobile base station. In Kloos, Carlos, Marn, Andrs, and Larrabeiti, David (editors): EUNICE 2005: Networks and Applications Towards a Ubiquitously Connected World, volume 196 of IFIP International Federation for Information Processing, pages 173–186. Springer Boston, 2006.

[42] Vergados, D.D.: Service personalisation for assisted living of elderly people at home. In Semantic Media Adaptation and Personalization, 2008. SMAP '08. Third International Workshop on, pages 135 –140, Dec. 2008.

[43] Walko, J.: Home control. Computing Control Engineering Journal, 17(5):16 –19, 2006, ISSN 0956-3385.

[44] Wang, Chia Chi, Chiang, Chin Yen, Lin, Po Yen, Chou, Yi Chieh, Kuo, I Ting, Huang, Chih Ning, and Chan, Chia Tai: Development of a fall detecting system for the elderly residents. In Bioinformatics and Biomedical Engineering, 2008. ICBBE 2008. The 2nd International Conference on, pages 1359 –1362, may 2008.

[45] Wang, S.: Intelligent Buildings and Building Automation. Taylor & Francis, London, 2010.

[46] Want, R., Farkas, K.I., and Narayanaswami, C.: Guest editors' introduction: Energy harvesting and conservation. Pervasive Computing, IEEE, 4(1):14 – 17, jan.- march 2005, ISSN 1536-1268.

[47] Werbos, P.J.: The roots of backpropagation: from ordered derivatives to neural networks and political forecasting. Adaptive and learning systems for signal processing, communications, and control. J. Wiley & Sons, Oxford, 1994, ISBN 9780471598978.

[48] Werbos, P.J.: Beyond regression: new tools for prediction and analysis in the behavioral sciences. Harvard University, 1974.

[49] Wikipedia: Squawk Virtual Machine. http://en.wikipedia.org/wiki/Squawk_virtual_machine (last access: 02.12.2010), 2010.

[50] Wong, A.C.W. and So, A.T.P.: Building automation in the 21st century. In Advances in Power System Control, Operation and Management, 1997. APSCOM-97. Fourth International Conference on (Conf. Publ. No. 450), volume 2, pages 819 –824 vol.2, November 1997.

[51] Yao, J., Simmons, S., and Warren, S.: Ease of use considerations for wearable point-of-care devices in home environments. In Distributed Diagnosis and Home Healthcare, 2006. D2H2. 1st Transdisciplinary Conference on, pages 8 –11, april 2006.

[52] Zensys: Z-wave Protocol Overview. An online version is available at http://www.smarthus.info/support/download/zwave/Z-Wave%20Protocol%20Overview.pdf (02.03.2011).

[53] Zhang, Xiaoyu, Jiang, Hanjun, Chen, Xinkai, Zhang, Lingwei, and Wang, Zhi- hua: An energy efficient implementation of on-demand mac protocol in medical wireless body sensor networks. In Circuits and Systems, 2009. ISCAS 2009. IEEE International Symposium on, pages 3094 –3097, may 2009.

[54] Zhou, Feng, Jiao, Jianxin, Chen, Songlin, and Zhang, Daqing: A case-driven am- bient intelligence system for elderly in-home assistance applications. Systems, Man, and Cybernetics, Part C: Applications and Reviews, IEEE Transactions on, 41(2):179 –189, march 2011, ISSN 1094-6977.

# 9 Appendix

## 9.1 Database Design

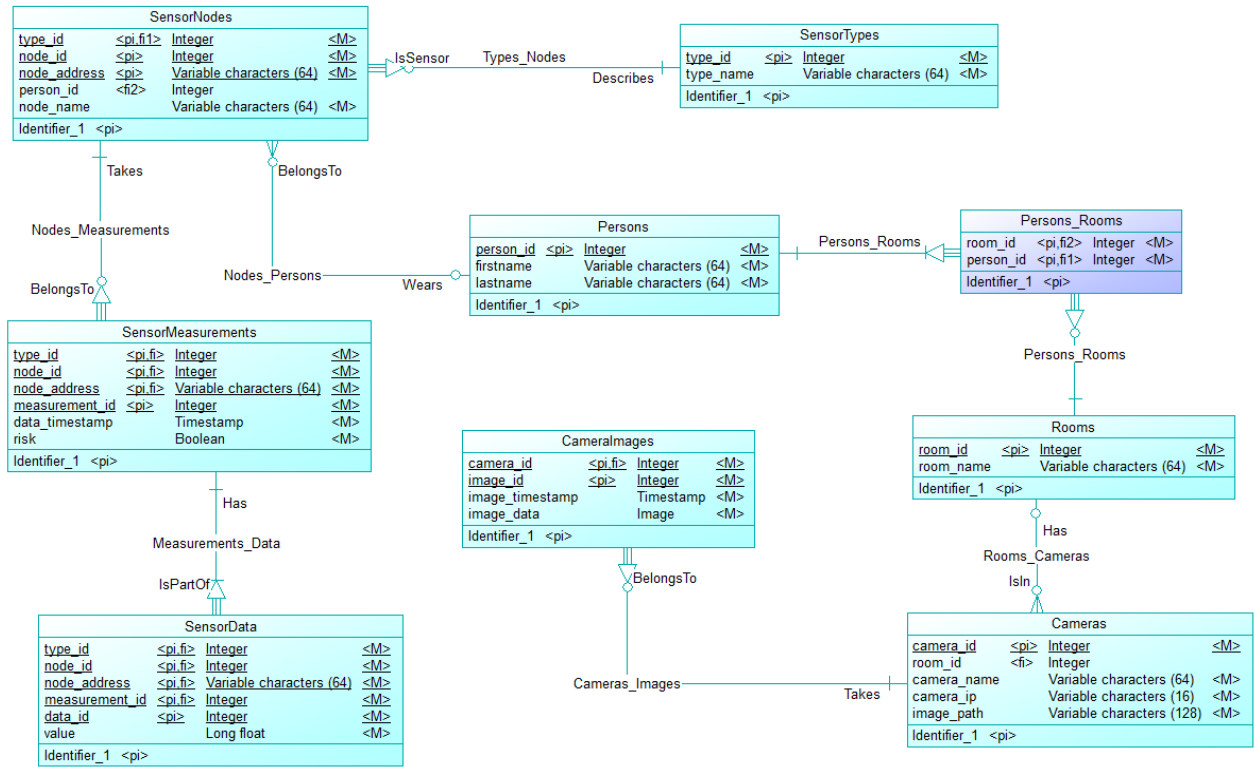

**Figure A. 1: Conceptual Data Model**

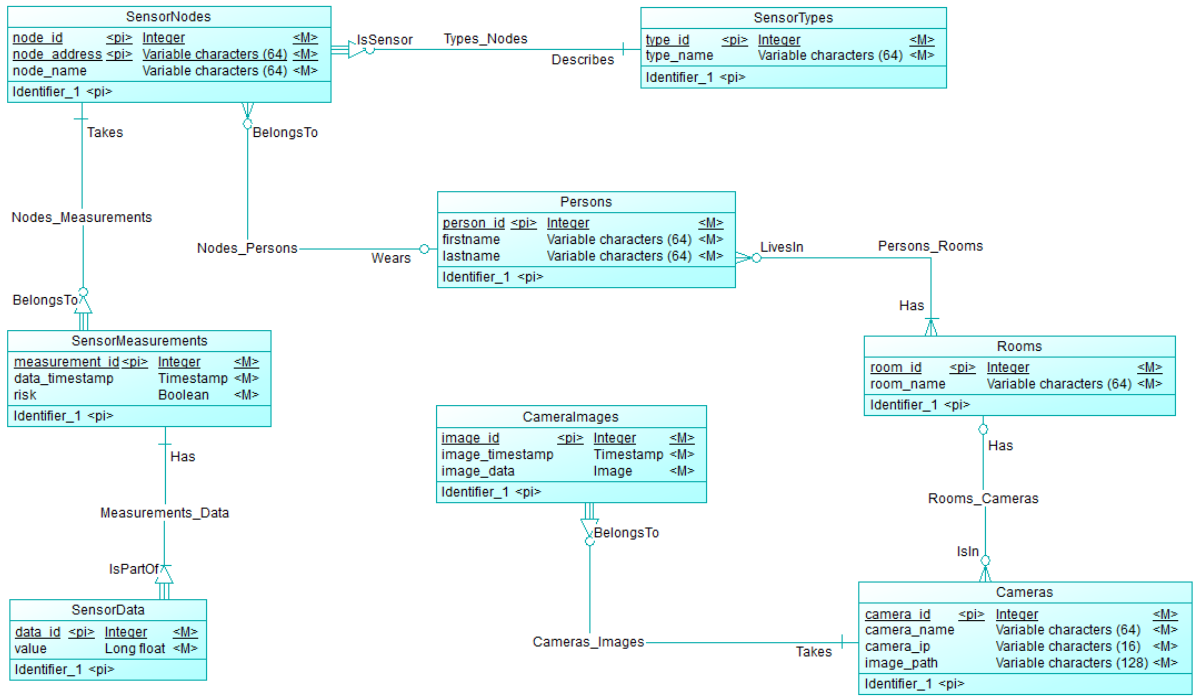

**Figure A. 2: Logical Data Model**

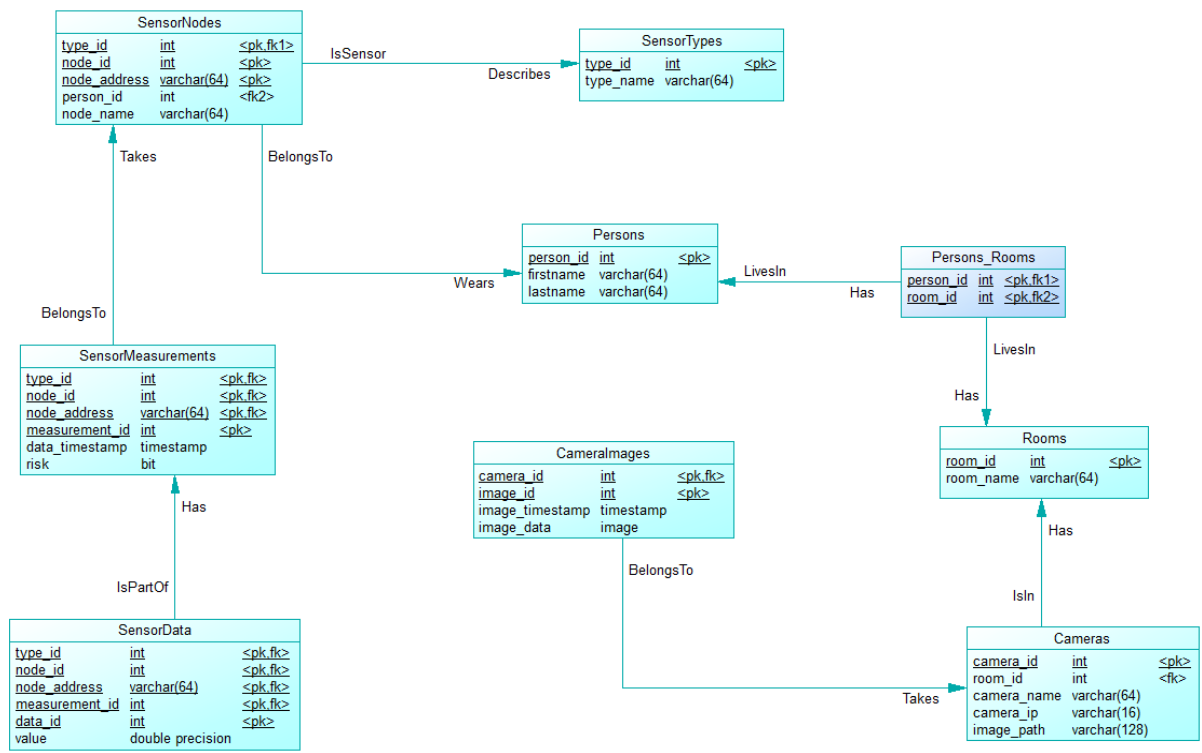

**Figure A. 3: Physical Data Model**

## 9.2 Historical Foundation of Elderly In-home Assistance

In order to provide background information about the origin of elderly in-home assistance, the authors of this thesis included this chapter of history of intelligent buildings (the ancestor of smart homes) and a definition of the term intelligent building itself. The following passage will provide a summed description of the development of building automation through the decades starting at the end of 1940s according to [50]. In addition to that the description is also depicting in a comprehensive way in figure 9-1 in order to guide the reader through this chapter.

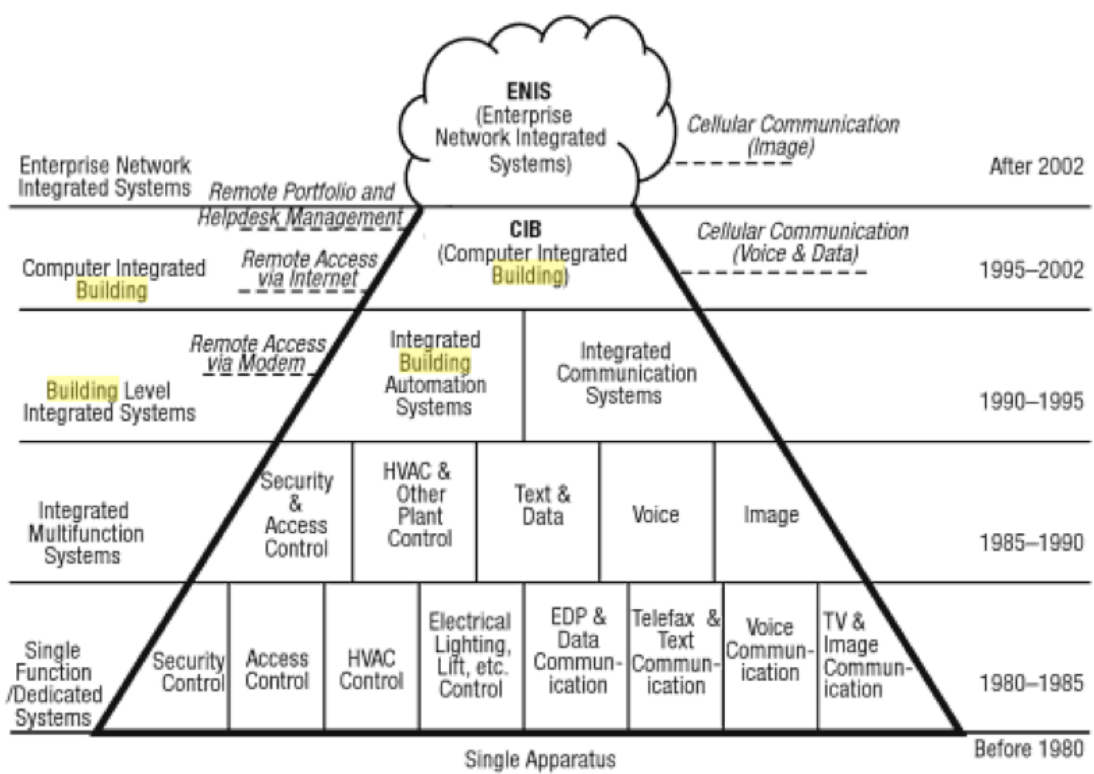

**Figure 9-1: Intelligent Pyramid, adapted from [45]**

One cannot fix the start of building automation or intelligent buildings (IB) at certain date. It is more like a flowing development starting in the years after World War II (WWII). The general expansion led to the demand of more comfort inside buildings which in return led to the development of more complex and better heating and cooling control systems. Pneumatic controls and electrical switches were used in order to achieve this task, still needed to be supervised by humans. In the 50s, pneumatic sensor transmitters together with possibilities of remote controlling triggered a paradigm shift from several controlling units to a more centralized approach by uniting all control operations into a single control room. This shift was supported by the on-going evolution of miniaturization, reducing the physical space requirements of the equipment more and more.

In the 60s, with help of control companies, electromechanical multiplexing systems were introduced to the market. This led to a huge decrease of costs for installation as well as maintenance. The multiplexer provided the technology in order to reduce the needed wiring inside the buildings. Based upon this development, there was no need for a whole building dedicated to control tasks anymore. Instead, all control functions could be combined into a single console, which included logging functions for the first time. These logging functions made it possible to automatically record out of norm values with their corresponding settings at this time in order to provide a baseline for an analysis afterwards.

From the beginning of the 70s, the new mini computers with central processing units (CPUs) and programmable logic controllers (PLCs) found their way into building automation. Together with the reactions to the oil embargo in 1973, energy management systems (EMS) became a standard and a lot of applications in that area came up like optimum start/stop programs or day/night control systems. As a side effect, also a lot of applications dedicated to security and fire alert popped up. With the new systems, owners of buildings were now able to compare the energy consumption of buildings and gain better cost control.

Finally, in the 80s the introduction of the personal computer (PC) revolutionized the control industry. Due to low chip prices, the speed of development in that area increased dramatically and the vendors of control systems also spent more money and time into developing new systems and not only tried to optimize present technologies. The so far used pneumatics was more and more replaced by distributed direct digital controls (DDDCs). Also, the control terminals were modified in order to provide more usability and user-friendly attributes, resulting in the exchange of old terminals to PC systems, which were meanwhile widely accepted due to their presence in universities and the industry. The programming also evolved to high level programming languages like C or Pascal.

So it is obvious that an intelligent building from the WWII is not the same any more than 30 to 40 years later, not to mention from today's state-of-the-art technology. Therefore, it is necessary to define the term intelligent building in order to fit today's needs and properties of current application areas. The upcoming chapter is discussing several definitions for intelligent buildings and is then defining a new, overall definition.

The definition of the term intelligent building varies a lot, depending on which organization is consulted or what country is chosen. According to Leifer (1988),

> *'An intelligent building is one in which the building fabric, space, services and information systems can respond in an efficient manner to the initial and changing demands of the owner, the occupier and the environment.' [7, p.6]*

This definition focuses on the integrative aspect of IBs meaning that every component, not only the technical ones, is contributing as subsystem of the building. Furthermore, the building's highest priority is to serve the owner, occupier and environment in an optimal and constant way.

The European Intelligent Building Group is concentrating mainly on the occupant of the IB, but adds another interesting aspect:

> *'[...] creates an environment which maximizes the effectiveness of the buildings occupants while at the same time enabling efficient management of*

*re- sources with minimum life-time costs of hardware and facilities.' [40, p.2]*

Thus, the tasks managed by the IB have to be accomplished in a way that uses as few resources as possible, while it protects used resources in terms of obtaining durability.

*'Use of technology and process to create a building that is safer and more productive for its occupants and more operationally efficient for its owners.' [5, p.17]*

This definition introduces the aspect of safety for its occupants. The term safer can be interpreted in two ways: the first one regarding the direct physical health of the occupant, while it also implies an indirect safety for the occupants being less harmful and therefore safer for the environment surrounding the IB.

*'An intelligent building would include a situation where the properties of the fabric vary according to the internal and external climates to provide the most efficient and user friendly operation in both energy and aesthetic terms. [3, p.39]*

Lush (1987) pushes with his definition above in a totally different direction. He points out the important fact of the IB's building materials to feature multiple properties in order to behave accordingly to the environment's influences. He also pinpoints the important aspect of aesthetics playing a major role regarding objects that are in daily use by humans.

An article in the New York Times (1984) rounds up this listing of definitions and provides an interesting metaphor to the topic of IB:

*'[...] an intelligent building has a computer for a brain and a fibre-optic- cable nerve system that tenants use for their telephone and data processing communications.' [29]*

This metaphor is interesting because it brings forth the aspects of human interaction. If the IB is seen as an abstract form of artificial being, parallels can be drawn to a situation of people living together. In order to 'live well together' the building has to communicate with the occupants via terminals as well as react accordingly to the actions of its tenants, recognized through its sensors.

For covering all the important aspects mentioned before, the authors came up with the following definition:

*An intelligent building is continuously adapting to its occupant, owner and environment regarding its material properties, as well as its services. This should be done on the one hand in a safe way while on the other hand aiming at a low demand of resources, combined with increased productivity and appealing aesthetics.*

## 9.3   Wireless  Technologies in Elderly In-home Assistance

This chapter is dedicated the important aspect of security in smart homes and in-home assistance setups. Therefore, three technologies are presented, addressed to provide security as well as energy efficient properties.

### 9.3.1 ZigBee

ZigBee is settled in the area of wireless networking technologies and was developed by the ZigBee Alliance[14] based on the IEEE 802.15.4 standard[15]. It is intended to provide connectivity for applications with characteristics of low-data rates and short distance transmissions. The protocol stack included in the ZigBee draft is based on four layers which can be seen in figure 9-2.

The Physical Layer (PHY) and the Medium Access Control Layer (MAC) are taken out of the IEEE 802.15.4 standard, while all layers above are elements of the ZigBee specification released by the ZigBee Alliance. For the PHY layer three frequency domains are used as there are 868MHz, 915MHz and 2.4GHz in Europe, the US and as a universal solution respectively. In order to access the channel, two main methods can be identified. The first is the Carrier Sense Multiple Access - Collision Avoidance (CSMA-CA) mechanism which is competitive and the second one which is based upon guaranteed time slots (GTS), being non-competitive.

Sitting on top of the MAC layer, the Network Layer (NWK) is responsible for the networks topology as well as the address assignment regarding the nodes inside the network. There are two different kinds of addresses. The first one has a length of 16 bits and is similar to the 'normal' IP addresses used on the Internet for example. The second kind is 64 bits long and can be compared to a MAC address of a network card. The first one is used by coordinator and routing devices inside the ZigBee network to assign addresses to end devices and use them for routing and transmission purposes [36].

---

[14]  http://www.ZigBee.org/
[15]  http://standards.ieee.org/getieee802/802.15.html

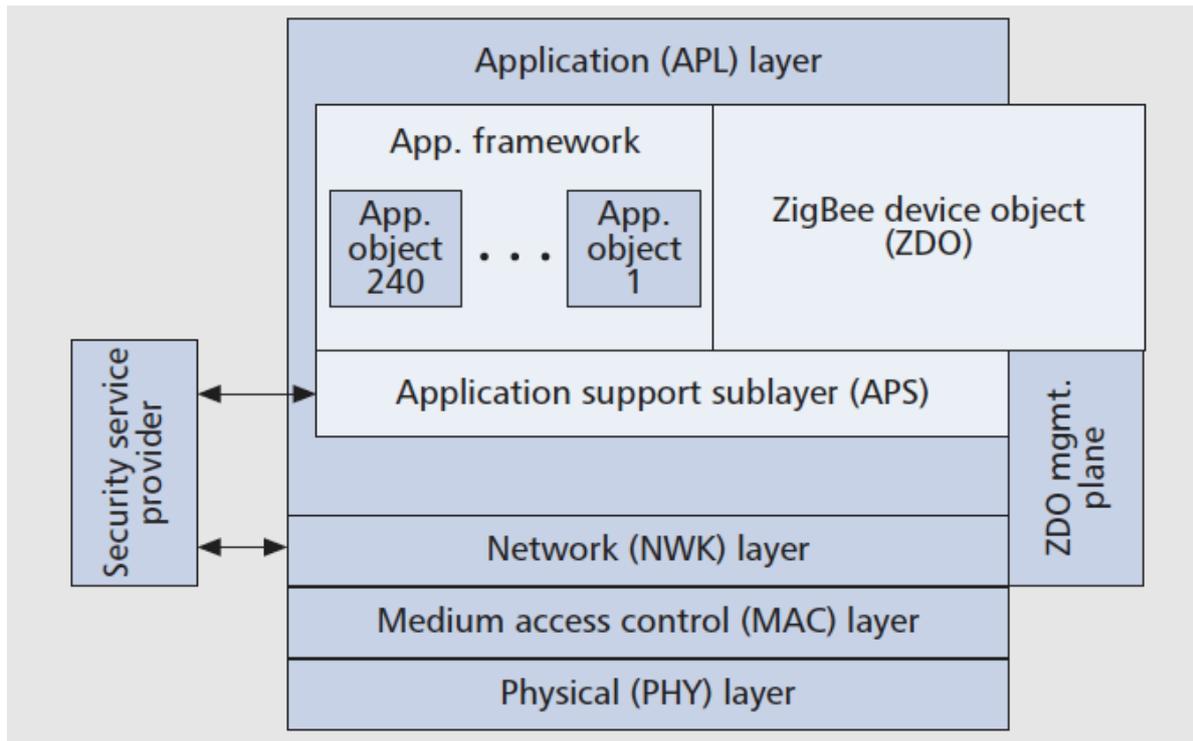

**Figure 9-2: ZigBee Protocol Stack, adapted from [13]**

The **Application Layer (APL)** represents the top level of the ZigBee stack. So called application objects reside inside this layer, developed by vendors in order to provide a customized device fitting best for the respective application. The whole stack management is done via these objects whose number can be up to 240 for each device. The ZigBee standard is also designed to provide profiles in order to support development of application objects as well as increase interoperability. These profiles are discussed more detailed in the upcoming section [12].

#### 9.3.1.1 ZigBee Profiles

The ZigBee technology includes several profiles[16], each of them dedicated to a special application area like:

- Home Automation
- Building Automation
- Smart Energy
- Telecommunication Services
- Health Care

Another example mentioned above is the **Home Automation Public Application Profile**. It includes recommendations for home-networks up to 500 nodes. This limit is not absolute, but experience shows that networks with a higher density of nodes are not working as reliably as within this given density. The target groups for this kind of profile are mainly private persons and professionals working within the area of home automation but without specific knowledge concerning ZigBee technology. The installation concepts included in this profile aim at easy installation and providing possibilities of cross-over Original Equipment Manufacturer (OEM) vendors integrations.

---

[16] http://www.zigbee.org/Standards/Overview.as

The installation shall provide a basic and sporadic real-time control of devices within the network. These real-time actions are triggered by the user by pushing buttons or activating switches in the house. The result shall end in a quick response to the user's actions with the desired outcome. During the remaining time, the network usually remains in an idle state. The network can contain either components of the standard feature set or the PRO feature set, as long as a ratio of about 50:50 is guaranteed. The more preferred solution however is a homogeneous environment of one feature set only. All used devices must be MSP certified in order to work properly. In addition to that, 'E-Mode' commission must be supported, defined by a 'one button push / two button push' approach or a remote controlling device of an OEM vendor.

The HAPA profile should not be mixed up with other ZigBee profiles due to possible incompatibility which occur due to security implementations in higher stack levels for example. Typical applications scenarios would be a 'on work' mode where all non-used electrical devices, including light and air-condition, are turned off or set to a minimum. Another scenario would be a home theatre where a single button dims the light, lowers the blinds and starts TV and DVD playback.

### 9.3.1.2 Security Considerations

Despite the good properties of ZigBee, one has to be aware of some security issues that have to be considered when dealing with ZigBee technology.

One of them is related to the Smart Energy Profile mentioned in the section before. According to Gianluca Dini and Marco Tiloca in [8] there exist concerns regarding the network- and link keys used in this profile. If a ZigBee device ends its current mission, it leaves the network as well as it does during maintenance. Of course, this means that it may not be able to connect again without proper reauthorization, so abuse is impossible. However, this implies that a proper key revocation / redistribution policy is established. This is not the case in the Smart Energy Profile. The easiest method would be that the Trust Centre simply deletes the key of the leaving device, so it is not valid anymore. But nothing in the profile description refers to how to handle leaving devices. This means that by compromising a device that has left the network and make it re-join, eavesdropping and the use of application commandos is possible, as well as injecting counterfeited information regarding routing or payload. The ZigBee specification indeed includes a statement about periodic updates, but either it states something about how to set it or grants the possibility of asynchronous triggers in case of a device leaving the network.

So the renewal of the networking key is the only possibility. Therefore, two methods are described in the ZigBee specifications, namely a broadcast-based and a unicast- based approach. The first one uses the old network key in order to protect the new one.

So this approach is not appropriate due to the compromised network key as described above already. The second approach uses a point-to-point method in order to distribute the new network keys, protecting them via a transport key. But this results in additional overhead and is only practical for a small amount of nodes due to scaling factors in encryption.

But not only is the network key endangered, also the link key used to connected devices directly. There is no hint in the specification what to do in case a device linked to another leaves the network. There are methods described regarding preparations to do if a device is signalling that it will leave, but this only aims on routing issues, not the danger of a compromised device hacking into another one via an old link key.

### 9.3.2 LonWorks

The Californian Company Echelon started to develop LonWorks and its underlying Control Network Protocol (ISO/IEC 14908-1) in 1988 [10, p.15]. The goal was to develop a cost-effective standard method for control networks using inexpensive control devices to communicate over a network by providing a standard, the devices of multiple manufacturers are able to cooperate. A complete platform has been developed which provides tools and methods for designing, creating and installing intelligent control devices.

Control Networks are mostly used in factory automation, building controls and embedded machines. LonWorks uses decentralized control network architecture [10, p.6] where the control over the network is intelligently distributed over the network. Intelligent control devices use a common protocol to communicate with each other and each device has embedded intelligence and control functions. Devices can range from simple sensors to complex supervisory controllers or data acquisition devices. Control networks are characterized by their small massages with low overhead which are transmitted frequently and need a high reliability to arrive at their target.

LonWorks is an open solution for building automation in industrial areas and home automation as well as in transportation and public utility control networks [10, p.7]. Due to the used decentralized architecture, processing is distributed in the whole network. Therefore, it is more robust and reliable as single points of failure are minimized. Using inexpensive devices and the open network control protocol allows lower installation and life-cycle costs and makes the network flexible to adaptions and extensions. The LonWorks platform consists among other things of the following components [10, p.16-18]:

- Smart Transceivers - consist of a neuron core and a transceiver. The neuron core is the processing unit and is named after the neurons in the brain because the control network works in a similar way. The Neuron Core implements Layer two to six of the ISO/OSI Reference Model, the device manufacturer adds application layer code to support the functionality of the device.

- Development Tools are used by the manufactures to develop applications for their produced devices.

- Routers are used as network components. They support multiple

transmission media and the creator of the network can decide which medium suits best for the needed application. The routers control the network traffic and can partition sections.

- Smart controllers - are used as web servers. It is possible to remotely access them and access the data managed by the smart server.

To connect the different devices of a LonWorks control network, two different possibilities exist [10, p.18]. Smaller networks of up to 200 devices can install themselves using the LonWorks interoperability self-installation (ISI) protocol [10, p.21]. There- fore, devices discover and communicate with each other, they configure themselves and each device takes over a part of the network management. The second option is the managed network in which device configuration and communication is defined by a user beforehand [10, p.20]. Managed networks also offer the possibility of monitoring services. The Management Server is only needed during the installation process and maintenance.

The assumptions behind the development of LonWorks have been that networked control system have a common requirement, independent of the area of application and that networked system are much more powerful than not connected ones [10, p.8]. The information in the packets sent through the control network is the same in different applications. Control Networks are optimized for performance, packet size and response requirements, therefore data networking technology is not appropriate [10, p.9].

The implementation of a control network requires wiring. Different technologies are supported to offer a solution for different areas of application. In factory automation twisted pair wires can be used, for home automation or vehicles like trains using the power lines of the power grid could be easier and therefore also a cheaper solution [10, p.10]. An effective system design is important. The processing capability which is defined by the channel capacity of a network can be increased by partitioning a large network into multiple smaller ones [10, p.11]. To make it possible for manufacturers to concentrate on the device itself, standard network services are provided. They do not need to develop a control system, and therefore they can use the development time won by improving the device.

### 9.3.2.1 Communication Layer Protocol

The LonWork implementation of the Control Network Protocol (ISO/IEC 14908-1) is called LonTalk [10, p.26]. It is optimized for control applications and provides a complete seven layer communication protocol.

Layer 1: The physical layer defines the transmission using a physical medium, called channel [10, p.31]. LonTalk supports multiple physical media like for example twisted pair cable, link power, radio, fibre optics or infrared. Routers can be used to connect multiple physical media. All supported media are bidirectional and the transmission rate depends on the medium and the transceiver used. A free network topology is supported [10, p.32] which allows any combination of star, loop and bus topologies. This has advantages, for example existing wiring in a building can be used for a control network and rewiring is only needed for extensions [10, p.34].

This reduces the time needed and costs of the installation or extension of a control network. Furthermore, a link power system is possible which allows the transmission of data and power on the same twisted pair cable. This can also be combined with locally powered devices. Another easy solution is the usage of the power line of a building. This is mostly used in home and transportation systems [10, p.39]. It uses the existing power line wiring as a data medium for the control network.

Layer 2: The link layer defines media access methods and data encoding [10, p.41]. If a device has a packet to send, it waits until the channel it needs to use is idle. Once it is idle, it waits a random interval and, if the channel is still idle, sends its packet. A unique characteristic of the control network protocol is that the number of randomizing slots increases with the increase in network load.

The media access control algorithm is based on the CSMA algorithm. P-persistent CSMA means that a device sends in a given randomizing slot with the probability P [10, p.42], Ethernet for example uses p is one [10, p.43]. This algorithm is dependent on collision detection, though this method is mostly not very reliable on many media used by control networks. The control network protocol uses a method called predictive p-persistent CSMA [10, p.44]. P is dynamically adapted to the current network load.

This is done by counting the number of devices which are expected to send data in the next cycle. Therefore, few packets sent and a light network load means that the number of randomized slots is smaller and the media access delays are minimized. During heavy network load a high number of randomizing slots is used and collisions can be minimized. Furthermore, the control network protocol supports a priority mechanism which allows reducing the response time of critical packets [10, p.44].

With a network management tool, a number of priority slots can be assigned. One device or a group of devices can be designated to a certain priority slot. Priority packets are added to a priority queue in a device and will be sent before any other non-priority packet. In routers, these priority packets are put at the front of the sending queue and the router will send the packages as priority packets if a priority slot has been assigned to the router.

Layer 3: The network layer defines the naming and addressing of devices and how packets are routed [10, p.50]. The name of each Neuron Core is a unique Neuron ID which is a 48bit number. It uniquely identifies the Neuron Core and does not change. The address is a unique identifier within a network. It is used by routers to transmit packets. It can be changed as needed. The Neuron ID could be used as address, but then only point-to-point connections would be allowed. The control network protocol uses a hierarchical addressing method [10, p.51].

The address consists of three parts, the domain which is used to communicate with domain components, the subnet which addresses a collection of devices and the node address which is unique for each device in a network. The domain is a logical collection of devices and network components which can use multiple channels. Also it is possible to have multiple domains on one channel to avoid interference between different networks using the same channel. The domain id is 0, 1, 3 or 6 bytes, it will be sent with each packet. Therefore, unnecessary overhead could be produced by choosing a long domain id if it is not needed. A maximum of 256 subnets are allowed per domain [10, p.52]. Subnets are logical collections of up to 127 devices.

These devices need to be on the same physical channel. The node id is a unique 7 bit code within the subnet it belongs to. Overall, a maximum of 32,385 devices are possible within one domain, but a control network may consist of multiple domains. Another possibility for a logical collection is the group [10, p.54]. Groups are within a domain and are independent of their physical location. They can be used very efficiently to send a packet to more than one device without addressing each device on its own.

Layer 4: The transport layer is responsible to ensure reliable delivery of messages [10, p.59]. There exist four basic types to ensure this [10, p.60].

- Acknowledged - If a receiver gets a packet addressed to it, it will send an acknowledgment back to the sender to confirm the packet. It is the most reliable method.

- Repeated - The sender sends a packet multiple times without expecting a response from the receivers. This method is used mostly for broadcasts to reduce the network load caused by the acknowledgments.

- Unacknowledged - This method is used to transmit packets whose loss would have no perceptible effect on the system, for example a packet from a periodic heartbeat.

- Request/Response - is handled by the session layer. It is similar to acknowledged, but each device sends a response with useful data, not only an acknowledgment that it received the packet.

Layer 5: The session layer controls the data exchange and implements an authentication protocol to ensure a sender is authorized to send a command [10, p.61]. For authentication, a 48bit key is used. The key is distributed for a domain during the installation process. To authenticate a message, the receiver challenges the sender with a random number. The number is transformed using the key. If sender and receiver transform the random number to the same result, the authentication is successful. This authentication is always enabled, if the key has not been set during installation a default key is used.

Layer 6: The presentation layer adds structure to the sent data and defines how messages are encoded [10, p.69].

Layer 7: The application layer provides a set of standard network services to configure and observe the network [10, p.79]. Through additional application standards it is ensured that the applications and devices from multiple manufacturers are able to communicate and cooperate with each other.

LonWorks for example has been used to control the street lightning in Oslo to reduce the streetlight operating costs, increase safety and allow the remote control and monitoring [11, p.1]. The devices communicate using the existing power line connections. Different sensors, for example weather sensors and internal astronomical clocks are used to determine the natural light of the sun and the moon to dynamically dim the streetlights. This allows a reduction in the energy costs and furthermore increases the lamp life span.

According to the company Real Time Automation[17], there are some drawbacks related to LonWorks when it comes to the topic security. In comparison to other technologies, LonWorks does not implement data encryption and is relying on sender authentication. The main difference lies in the focus of the techniques. While data encryption is used to make data unreadable for third persons, sender authentication proves that the sender's provided identity is its real one and, one step further, whether the sender then is allowed to send or not.

In order to prove, the sending device's authentication request, a 64bit random number is send to the device. The receiving device transforms this number with its own, secret and unique ID number and sends back the result. Its counterpart, knowing the secret ID of all devices that may contact it, does the same operation and then compares the returned result from the other device with its own transformation result. If they match, the requesting device has been authenticated successfully.

However, this method only guarantees that devices only accept data from authenticated devices. It does not protect the data in the network from being eavesdropped, nor does it prevent the possibility of hacking a central device and therefore gaining all the stored keys. Also, there is the potential danger of Man-in-the-Middle Attacks in order to perform a reply attack to the central device. Doing so, the transformed answer message could be intercepted and used for own transmission purposes.

### 9.3.3 Z-Wave

Z-Wave is a proprietary wireless communication standard developed and maintained by the Danish company Zensys and the Z-Wave Alliance. The standard was developed to fit special needs in building and home automation. The Z-Wave alliance is a coalition of independent manufacturers which are specialized in wireless home automation systems. With about 160 alliance members and 170 products, Z-Wave is market leader in the 'wireless home control' sector. Well-known members of the alliance are Danfoss, Intel Corporation, Monster Cable, Universal Electronics, Wayne-Dalton and Zensys itself. The Z-Wave standard is a direct competitor of the ZigBee standard[18].

Z-Wave provides a transmission rate between 9600 bits per second and 40000 bits per second. It uses 2-FSK (frequency shift keying) for signal transmission. This technique modulates a carrier frequency with the data signal. The modulated signal is sent and demodulated by the receiver. This transmission technology provides high robustness to noise signals. Z-Wave uses the industrial, scientific and medical (ISM) radio band which is located at around 900MHz in the frequency spectrum.

In Europe, the used frequency is 868.42MHz and in USA it is 908.42MHz. The transmission range for Z-Wave communication is about 200 meters in the open field. Depending on building materials, the transmission range indoors is 30 meters maximum.

The Z-Wave standard uses a meshed network topology which means that each device,

---

[17] http://www.rtaautomation.com/LonWorks/
[18] http://www.z-wavealliance.org/modules/AllianceStart/

also called node, is connected to one or more other nodes in the network. The advantage of meshed networks is the fact that messages between two nodes can be transmitted although these nodes are not directly connected with each other, e.g. the distance between these two nodes is too high. The message is then transmitted using other nodes in between.

Thus, a Z-Wave network can span much further than the range of a single device. This type of transmission is called routing. Z-Wave enables the use of 232 devices in a single network. Furthermore, it is possible to connect two or more networks with each other by installing so called bridges[19] [19, 43].

The Z-Wave protocol provides interoperability for all products which means that there is a standardized communication across any Z-Wave product. The interoperability is achieved by standardizing on two levels: command level and device level. In the command level, all transferable commands between nodes are standardized. In the device level, all products must be a member of a certain device class which defines mandatory, recommended and optional commands and their parameters[20]

### 9.3.3.1 Protocol Stack

The Z-Wave protocol stack consists of 4 layers:

- Physical Layer
- Transport Layer
- Network/Routing Layer
- Application Layer

The physical layer, also called PHY/MAC layer, controls and gives access to the radio frequency (RF) media. As mentioned before FSK with Manchester channel encoding is used in the PHY/MAC layer. The data stream consists of a preamble, start of frame (SOF), frame data and end of frame (EOF) symbols. The frame data is the only part which is passed onto the transport layer. In general, the PHY/MAC layer is independent of the RF media, frequency and modulation method used for transmission. Further, the PHY/MAC layer has a collision avoidance mechanism included which prevents nodes from starting a transmission while other nodes have active transmissions [52, 28].

The transfer layer is responsible for data transmission between two network nodes. This transmission also includes acknowledgements for confirming successful transmission, checksum checks for detecting transmission errors and retransmission in cases of unacknowledged data packages or detected checksum errors. The transfer layer uses different frame layouts for each task mentioned above. The used frames are single- cast frames, transfer acknowledgement frames, multicast frames and broadcast frames [52, 28]. The network/routing layer provides correct package routing between two network nodes as well as data retransmission. Further, it scans the network topology in order to build routing tables and perform routing table updates. Each node in a Z-Wave network has routing capabilities. This is the key feature of the Z-Wave standard. The routing layer provides two frames: a routed single cast frame which is a one-node destination frame which contains repeater information and a routed acknowledge frame which is a routed single-cast frame without payload to confirm that a routed single cast frame has reached its destination. Routing tables stores information about the network topology. They are bit field tables which store connection between nodes in the mesh network [52, 28].

---

[19] http://www.z-wavealliance.org/modules/AllianceStart/
[20] http://www.hometoys.com/htinews/jun05/articles/zensys/homecontrol.htm

The application layer covers all application, specific commands as well as the Z-Wave specific commands for a node. It decodes and executes commands in a Z-Wave network. This is done in a main loop. Each command is defined within a command class. Each command may contain any parameters associated with the specific command. The number of available parameters depends on the command itself. A large part of the application layer, despite the assignment of Home IDs and Node IDs, is implementation- specific and can be different in each implementation [52, 28].

### 9.3.3.2 Security Considerations

The original version of the Z-Wave standard used a security algorithm called triple data encryption algorithm (DES) with a key of 56 bits. Due to the fact that Z-Wave customers claimed that security was not an issue, the security feature and security layer were removed from the Z-Wave protocol. Therefore, the Z-Wave standard lacks data encryption and transmits network information in plain text which allows eavesdroppers to capture this information. It is then possible to determine which command controls devices functionality.

Eavesdroppers could rebroadcast the command to force certain actions, e.g. to open a door or turn lights and intruder detections on and off. The only way to protect sensible devices from unauthorized access in Z-Wave networks is to use a so called rolling code which is already used in automated door locks for cars. A transmitter sends an access code, with function instructions. The receiver stores the current access code in memory. If it receives the same access code again the instruction is accepted. Transmitter and receiver must use the same pseudo-random number generator to pick identical new codes when an access code is sent or received which means that transmitter and receiver must be synchronized. This method is only vulnerable to hacks done by using powerful computers.

Future Z-Wave standards may include the advanced encryption standard (AES) to provide powerful security functionalities. This is required for the use in non-residence and commercial buildings. Z-Wave has a special procedure for setting up new devices in an existing Z-Wave network. The initial transmissions are run at low power which makes them undetectable at some distance away. Furthermore, buttons on the controller as well as buttons on the device must be pressed simultaneously [19, 43].

### 9.3.4 Comparison between ZigBee and Z-Wave

Since LonWork is focused differently than ZigBee and Z-Wave, the authors decided to give a short, point-wise overview of the differences between Z-Wave and ZigBee:

- ZigBee is a non-proprietary standard, while Z-Wave is a proprietary standard
- Both are wireless standard
- Both provide the possibility of meshed networks
- Both are designed for low power consumption
- ZigBee provides higher data rate than Z-Wave
- ZigBee has built-in security features, which Z-Wave is lacking
- Z-Wave focuses on residential command, ZigBee on industrial, residential, and medical sectors
- Z-Wave modules are half the size of ZigBee modules due to lack of security
- Z-Wave suffers of higher latency than ZigBee

Due to its lower latency, its focus on the medical sector with its profiles, security features and non-proprietary properties, ZigBee will be used for the prototype and simulations. This also goes along very well with the fact that the radio module used in the testing devices is based on IEEE 802.15.4 which is the basis for ZigBee as well.